\documentclass[prb,twocolumn,showpacs]{revtex4}
\usepackage{bm}
\usepackage{graphicx}

\begin{document}
\title{The spherical $2+p$ spin glass model:  an analytically solvable model
     with a glass-to-glass transition.}

\author{A. Crisanti}
\email{andrea.crisanti@roma1.infn.it}

\author{L. Leuzzi}
\email{luca.leuzzi@roma1.infn.it}
\affiliation{Dipartimento di Fisica, Universit\`a di Roma ``La Sapienza''}
\affiliation{Istituto Nazionale Fisica della Materia, Unit\`a di Roma, 
             and SMC,
             P.le Aldo Moro 2, I-00185 Roma, Italy}
\affiliation{Istituto Studi della Complessit\'a (ISC), CNR, 
             Via dei Taurini 19, I-00185 Roma, Italy}

\begin{abstract}
We present the detailed analysis of
the spherical $s+p$ spin glass model with two competing interactions:
among $p$ spins and among $s$ spins.  The most interesting case is the
$2+p$ model with $p\geq 4$ for which a very rich phase diagram occurs,
including, next to the paramagnetic and the glassy phase represented
by the one step replica symmetry breaking {\em ansatz} typical of the
spherical $p$-spin model, other two amorphous phases.  Transitions
between two contiguous phases can also be of different kind.  The
model can thus serve as mean-field representation of
amorphous-amorphous transitions (or transitions between undercooled liquids of
different structure).  The model is analytically solvable everywhere
in the phase space, even in the limit where the infinite replica
symmetry breaking {\em ansatz} is required to yield 
a thermodynamically stable phase.

\end{abstract} 


\maketitle


Spin glasses have become in the last thirty years the source of ideas
and techniques now representing a valuable theoretical background for
``complex systems'', with applications not only to the physics of
amorphous materials, but also to optimization and assignment problems
in computer science, to biology, ethology, economy and finance.  These
systems are characterized by a strong dependence from the details,
such that their behavior cannot be rebuilt starting from the analysis
of a 'cell' constituent but an approach involving the collective
behavior of the whole system becomes necessary. One of the feature
usually expressed is the existence of a large number of stable and
metastable states, or, in other words,  a large choice in
the possible realizations of the system and a rather difficult (and
therefore slow) evolution through many, details-dependent,
intermediate steps, hunting its equilibrium state or optimal
solution.

Mean-field models have largely helped in comprehending many of the
mechanisms yielding such complicated structure and also have produced
new theories (or combined among each other old concepts pertaining to
other fields) such as, e.g., the spontaneous breaking of the replica symmetry 
and the ultrametric structure of states.

Among mean-field models
 spherical models are analytically solvable even in the
most complicated cases.  Up to now only spherical models with one step
Replica Symmetry Breaking (1RSB) phases were studied, mainly due to
their relevance for the fragile glass transition.
\cite{KirThi87,CriSom92,REV} The possibility of the existence of 
Full Replica Symmetry Breaking (FRSB) phases in spherical models was
first pointed out by Nieuwenhuizen \cite{Nieuwenhuizen95} on the basis
of the similarity between the replica free energy of some spherical
models with multi-spin interactions and the relevant part of the free
energy of the Sherrington-Kirkpatrick (SK) model.
\cite{BraMor78,PytRud79}  A complete analysis, however, was not
provided up to now.  The problem has been considered some years later
\cite{CiuCri00} in connection with the possible different scenarios
for the critical dynamics near the glass transition, \cite{GoeSjo89}
therefore analyzing only the dynamical behavior in the 1RSB phase.

The model we present here, the $2+p$ spherical spin glass model,
displays four different phases: together with the replica symmetric,
the 1RSB and the FRSB phases also a new  phase occurs.  The
evidence for the existence of such peculiar amorphous phase has been
first presented in Ref. \onlinecite{CriLeu04} and, for what concerns
the organization of the states, it seems to yield the properties of a
glass up to the first level of the ultrametric tree (i.e. inside and
just outside a valley of the free energy landscape) and those of a
spin-glass above.

Concentrating on the study of amorphous materials, in recent years
some evidence has been collected for the existence of amorphous to
amorphous transition (AAT), in certain glass-forming substances.  One
way of looking at an AAT has been  to consider the kinetics of the
coordination transformation occurring in  strong glasses such as the
vitreous Germania (GeO$_2$ - from fourfold to sixfold coordination
rising the pressure) and Silica (SiO$_2$ - from tetrahedral to octahedral
coordination).\cite{Tsiok98}
Exactly as for the liquid-glass transition also this
transition is not a thermodynamic one, but it amounts to a qualitative
change of the (slow) relaxation dynamics, apparently expressing a
recombination of the glass structure (see also the numerical
simulations of Ref. \onlinecite{Huang04} for a different point of view).
Another kind of pressure induced AAT takes place in densified porous
silicon, where the high-density amorphous Si transforms into a low
density amorphous Si upon decompression.\cite{Deb01} A similar
transition also takes place in undercooled water.\cite{Poole}

Theoretical models have been introduced to describe an AAT. As, for
instance, a model of hard-core repulsive colloidal particle subject to
a short-range attractive potential that induces the particle to stick
to each other. \cite{DawetAl01,Zac01,Sciorti02} In the framework of
the Mode Coupling Theory (MCT) it has been shown that the interplay of
the attractive and repulsive mechanisms results in the existence of a
high(er) temperature ``repulsive'' glass, where the hard-core
repulsion is responsible for the freezing in of many degrees of
freedom and the kinetic arrest, and a low(er) temperature
``attractive'' glass that is energetically more favored than the other
one but only occurs when the thermal excitation of the particles is
rather small. Such theoretical and numerical predictions seem to have
been successfully tested in recent experiments.\cite{Chen03, Eckert02,
Pham02} Another model where AAT is found is the spherical $p$-spin
model on lattice gas of Caiazzo {\em et al.}  \cite{Caiazzo04} where
an off-equilibrium Langevin dynamics is considered, thus going beyond
the MCT assumption of equilibrium.
 The model we consider here might, as well, be a good mean-field
representative of an amorphous-amorphous transition.

The goal of this paper is to give a detailed discussion of the
different solutions  describing the low temperature phase of the
spherical $2+p$ spin glass model.  As it happens in systems with a
phase described by a 1RSB solution one must distinguish between the
``static solution'' obtained from the partition function and the
``dynamic solution'' obtained from the relaxation
dynamics.\cite{KirWol87,CriSom92,CriHorSom93} To keep the length of
the paper reasonable we shall consider in detail only the static
approach and introduce the the dynamic solution with the help of the
complexity.\cite{CriNie96} The complete dynamic approach will be
presented elsewhere. 

In section \ref{model} we present the spherical $2+p$ spin-glass
model. In section \ref{static} its static behavior is studied with the
help of the replica trick\cite{EA75} and the Parisi replica symmetry
breaking scheme:\cite{Parisi80} four different phases occur, together
with the relative transitions between them. The nature of the phases
is thoroughly discussed and analytical exact solutions for order
parameters, transition lines and thermodynamic functions are provided
all over the parameter space.  In section \ref{complexity} the
existence of an exponential number of energetically degenerate pure
states is considered by analyzing the complexity function. The
connection between the ``marginal condition'' (maximum of the
complexity in free energy) and the dynamical solution leads, in
section \ref{dynamics} to the discussion of the latter in those cases
where it differs from the static one.  In Appendix A and B we show,
respectively, the Parisi anti-parabolic equation for the $2+p$ model,
and its analytical solution. In appendix C some basic features of the
behavior of the much simpler $s+p$ model ($s$, $p>2$) are
given.\cite{CiuCri00,AnetAl} Eventually, in appendix D, a proof is given
that no phases other than those here presented exist for the spherical
$2+p$ spin glass model.

\section{The Model}
\label{model}
The spherical $2+p$ spin glass model is defined by the Hamiltonian
\begin{equation}
\label{eq:ham}
{\cal H} =  \sum_{i<j}^{1,N}J^{(2)}_{ij}\sigma_i\sigma_j
           +\sum_{i_1<\ldots <i_p}^{1,N}J^{(p)}_{i_1\ldots i_p}
           \sigma_{i_1}\cdots\sigma_{i_p}
\end{equation}
where $p$ is an integer equal or larger than $3$ and $\sigma_i$
are $N$ continuous real spin variables which range from $-\infty$ to 
$+\infty$ subject to the global spherical constraint
\begin{equation}
\label{eq:sphconst}
\sum_{i=1}^{N} \sigma_i^2 = N.
\end{equation}
The coupling strengths $J^{(p)}_{i_1\ldots i_p}$ ($p=2,3,\ldots$) 
are quenched independent identical distributed
zero mean Gaussian variables of variance
\begin{equation}
\label{varVp}
  \overline{\left(J^{(p)}_{i_1 i_2..i_p}\right)^2} = 
   \frac{p!\, J_p^2}{2\,N^{p-1}}, 
  \qquad i_1 < \cdots < i_p.
\end{equation}
The scaling with the system size $N$ ensures an extensive free energy and hence
a well defined thermodynamic limit $N\to\infty$. Without 
loosing in generality one may take either $J_2$ or $J_p$ equal to $1$
since this only amounts in a rescaling of the temperature $T$.
To keep the discussion as simple as possible in this paper we shall not 
consider the effect of an external field coupled linearly with the spin 
variables $\sigma_i$.

The properties of the model strongly depend on the value of $p$. 
For $p=3$ the model reduces to the usual spherical $p$-spin spin glass 
model in a field\cite{CriSom92} with a low temperature phase described by
a 1RSB solution. 
For $p>3$ the model exhibits different low temperature phases
which, depending on the temperature and the ratio $J_p/J_2$ between the 
strength of the non-harmonic and the harmonic parts of the Hamiltonian,
are described by 1RSB and/or FRSB solutions.


\section{The Static Solution}
\label{static}
The static solution is obtained from the minimum of the free-energy functional
computed from the partition function. 
The model contains quenched disorder and hence the partition function 
must be computed for fixed disorder realization:\cite{Ma76,MPV87,Fischer91}
\begin{equation}
   Z_N[{\bm J}^{(2)},\bm{J}^{(p)}]= \mbox{Tr}_{\sigma}\, 
        \exp\bigl(-\beta {\cal H}[\bm{J}^{(2)},\bm{J}^{(p)};\sigma]\bigr)
\end{equation}
with $\beta = 1/T$.
We have explicitly shown the dependence of the Hamiltonian on
the realization of the random couplings to stress that $Z_N$ is itself a 
function of the couplings realization.
The trace over the spins is defined as:\cite{CriSom92}
\begin{equation}
  \mbox{Tr}_{\sigma} \equiv 2\sqrt{N} \int_{-\infty}^{+\infty}\,
                            \prod_{i=1}^{N}\, d\sigma_i\, 
			    \delta\left(\sum_{i=1}^{N}\sigma_i^2 - N\right)\,
\end{equation}
and includes the spherical constraint (\ref{eq:sphconst}). As a consequence
$\mbox{Tr}_{\sigma}(1)$ is equal to the surface of the $N$ 
dimensional sphere of radius $N^{1/2}$ and its logarithm gives
the entropy of the model at infinite temperature.

The partition function $Z_N$ is a random variable, therefore the 
{\em quenched} free energy per spin is given by 
\begin{equation}
\label{eq:freeque}
\Phi_N = -{1\over N\beta}\ \overline{\ln Z_N[{\bm J}^{(2)},\bm{J}^{(p)}]}
\end{equation}
where here, and in the following, 
$\overline{(\cdots)}$ denotes the average over the realizations 
of all couplings in the Hamiltonian:
\begin{equation}
{\overline{(\cdots)}} 
=
\int
d{\cal P}[\bm{J}^{(2)}]\, d{\cal P}[\bm{J}^{(p)}]\,   (\cdots)
\end{equation}          

The thermodynamic limit $N\to\infty$ 
of the free energy $\Phi=\lim_{N \to \infty} \Phi_N$
is well defined and is equal to the limit 
$-\lim_{N\to\infty} \ln Z_N[\bm{J}^{(2)},\bm{J}^{(p)}]\,/\, N\beta$
for almost all coupling realizations (self-average property).

The analytic computation of the quenched free energy, i.e., of the average of
the logarithm of the partition function, is  quite a difficult problem,
even in simple cases as nearest neighbor one dimensional models.
Since the integer moments $\overline{Z_N^n}$ 
of the partition function are easier to compute, the standard method 
to evaluate (\ref{eq:freeque}) uses the so called ``replica trick''
by considering the {\em annealed} free energy $\Phi(n)$ of $n$ non-interacting
identical `replicas' of the system,\cite{SKPRL75,MPV87,Fischer91}
\begin{equation}
\Phi(n)= -\lim_{N\to\infty} {1\over N\beta n}\ \ln
                           \overline{(Z_N[\bm{J}^{(2)},{\bm J}^{(p)}])^n}.
\end{equation}
The quenched free energy $\Phi$ is then recovered as the
continuation of $\Phi(n)$ down to the unphysical limit $n=0$,\cite{note21}
\begin{equation}
   \Phi= -\lim_{N\to \infty}\lim_{n\to 0}
                   {\overline{(Z_N[J^{(2)},J^{(p)}])^n} - 1\over N\beta n}
      = \lim_{n\to 0} \Phi(n).
\label{TP}
\end{equation}
In the last equality we assumed that the replica limit $n\to 0$ 
and the thermodynamic limit $N\to\infty$ can be exchanged. 
The existence of such a limit 
has been recently rigorously proved.\cite{Talagrand, Guerra} 

The replica method gives a simple way of performing the disorder average,
at the expenses of introducing an effective interaction among different 
replicas in the $n$-dimensional replica space.
The interested reader can find a detailed 
presentation of the replica method for disordered
systems in Refs. [\onlinecite{MPV87}] and [\onlinecite{Fischer91}]
and for the particular case of spherical models in 
Ref. [\onlinecite{CriSom92}].

Applying the replica method the integer moments of the partition function
of the spherical $2+p$ spin glass model can be written, neglecting all
unnecessary constants and terms irrelevant for $N\to\infty$, as:\cite{CriSom92}
\begin{equation}
\label{eq:znav}
\overline{Z_N^n} = {\rm e}^{nNs(\infty)}\, \int_{\bm{q}> 0} 
            \prod_{\alpha<\beta}\,dq_{\alpha\beta}\
            {\rm e}^{NG[\bm{q}]}
\end{equation}
where $s(\infty) = (1+\ln 2\pi) / 2$ is the entropy per spin at infinite 
temperature and $G[\bm{q}]$ the functional:
\begin{eqnarray}
\label{eq:free2}
G[{\bm q}] &=& \frac{1}{2} \sum_{\alpha\beta}^{1,n} g(q_{\alpha\beta}) + 
       \frac{1}{2}\ln\det {\bm q}
\\
\label{eq:free3}
g(x) &=& \frac{\mu_2}{2} x^2 + \frac{\mu_p}{p} x^p.
\end{eqnarray}
with $\mu_p = (\beta J_p)^2 p / 2$.
We also introduce the two additional functions
\begin{equation}
\Lambda(x) = \frac{d}{dx}\, g(x), \quad
\Sigma(x) = \frac{d}{dx}\, \Lambda(x)
\end{equation}
whose utility will be clear in a short while.

The symmetric $n\times n$ real matrix $q_{\alpha\beta}$ is  the
replica overlap matrix 
\begin{equation}
\label{eq:qab}
q_{\alpha\beta} = \frac{1}{N}\sum_{i=1}^N\, \sigma_i^\alpha\,\sigma_i^{\beta},
\quad \alpha,\beta = 1, \ldots, n.
\end{equation}
The spherical constraint, Eq. (\ref{eq:sphconst}), implies that 
the diagonal elements of the matrix $\bm{q}$  are all equal to one:
$q_{\alpha\alpha} = \overline{q} = 1$.

In the thermodynamic limit $N\to\infty$ the integrals in (\ref{eq:znav})
can be evaluated by the saddle point method and the quenched free energy 
per spin $\Phi$ reads:
\begin{equation}
\label{q:free1}
-\beta \Phi = s(\infty) + 
           \lim_{n\to\ 0} \frac{1}{n} G[{\bm q}]
\label{eq:free1}
\end{equation}
where $G[\bm{q}]$ must be evaluated on the solution of the saddle point 
equation which, in the $n\to 0$ limit, reads:
\begin{equation}
\label{eq:spab}
\Lambda(q_{\alpha\beta}) + ({\bm q}^{-1})_{\alpha\beta} = 0, \qquad
\alpha\not=\beta.
\end{equation}

Stability of the saddle point calculation requires that the quadratic
form
\begin{equation}
\label{eq:stabil}
-\sum_{\alpha\beta}\,\Sigma(q_{\alpha\beta})\,(\delta q_{\alpha\beta})^2 + 
          \mbox{Tr}({\bm q}^{-1}\, \delta {\bm q})^2
\end{equation}
must be positive definite. Here above 
$\delta{\bm q}_{\alpha\beta}= \delta q_{\alpha\beta}\,
(=\delta q_{\beta\alpha})$ is the fluctuation of $q_{\alpha\beta}$ from the 
saddle point value (\ref{eq:spab}).
Details of 
the derivation of these equations can be found in
 Ref. [\onlinecite{CriSom92}].

The structure of the overlap matrix $q_{\alpha\beta}$ reflects the
organization of the different thermodynamic states, called {\em pure states},
in which each replica can be found. This however
does not follow from the replica calculation and therefore to evaluate 
explicitly $G[{\bm q}]$ some {\it ansatz} on the structure of $\bm{q}$ 
must be imposed. 


\subsection{The Replica Symmetric Solution (RS)}
The simplest {\it ansatz} is the one in which all
replicas are in the same pure state, so that 
$q_{\alpha\beta}$ cannot depend on the replica indexes:
\begin{equation}
\label{eq:qrs}
q_{\alpha\beta} = (1-q)\delta_{\alpha\beta} + q.
\end{equation}
This is called the Replica Symmetric (RS) {\it ansatz}.
This assumption is reasonable for coupling strengths
not too large or high temperatures, i.e., $\mu_2$ and $\mu_p$ 
small enough. 
In both cases, indeed, the system can explore almost the whole
available phase 
space so that different replicas will be found in the same pure state.

Inserting the RS form (\ref{eq:qrs}) of $q_{\alpha\beta}$ into
Eq. (\ref{eq:free2}) one gets
\begin{equation}
2 \lim_{n\to 0}\frac{1}{n}\,G[{\bm q}] = g(1) - g(q)
                          + \ln(1-q) + \frac{q}{1-q}
\end{equation}
where $q$ is the solution of the RS saddle point equation
\begin{equation}
\label{eq:rsq}
\Lambda(q) - \frac{q}{(1-q)^2} = 
\mu_2 q + \mu_p q^{p-1} - \frac{q}{(1-q)^2} = 0.
\end{equation}
In absence of external fields the saddle point equation
always admits the ``paramagnetic'' solution $q=0$. However
since $q/(1-q)^2$ diverges as $q\to 1$ and vanishes for $q=0$ 
for particularly chosen values of the parameters $\mu_p$ and $\mu_2$
there may also be solutions with $0< q < 1$.

The RS solution is stable, i.e. the quadratic form (\ref{eq:stabil}) is 
positive definite for $n\to\ 0$, provided that the eigenvalue\cite{note7}
\begin{eqnarray}
\label{eq:rs-stab}
\Lambda_1 &=& -\Sigma(q) + \frac{1}{(1-q)^2}  \nonumber\\
            &=& -\mu_2 - \mu_p (p-1) q^{p-2} + \frac{1}{(1-q)^2}
\end{eqnarray}
is positive.
If $q\not= 0$ then dividing the saddle point 
equation (\ref{eq:rsq}) by $q$ and adding the result 
to (\ref{eq:rs-stab}) one gets that the requirement $\Lambda_1>0$ 
is equivalent to
\begin{equation}
\label{eq:qunsta}
-\mu_p (p - 2)\, q^{p-2} > 0.
\end{equation}
This inequality cannot be satisfied for any $q>0$,
thus we are left with the $q=0$ solution only.

For $q=0$ the eigenvalue $\Lambda_1$ reduces to $\Lambda_1 = 1-\mu_2$. 
Therefore, in the $(\mu_p,\mu_2)$ plane 
the paramagnetic RS solution $q=0$ is stable everywhere below the $\mu_2=1$
line, which represents
the De Almeida-Thouless line\cite{AT} of the model.

The instability of the paramagnetic solution is due to the 
presence of the quadratic term in the Hamiltonian. If this is
missing, as for example in the spherical $3+p$ spin glass model in which the 
two-body interaction is replaced by a three-body interaction, the paramagnetic 
solution is stable everywhere in the phase space, similarly
to what happens for the spherical $p$-spin model without a 
field.\cite{CriSom92} 


\subsection{The One Step Replica Symmetry Breaking Solution (1RSB)}
The stability of the RS solution $q=0$ does not depend on $\mu_p$. However,
from the analogies with the spherical $p$-spin spin glass model we expect that 
for $\mu_p$ large enough a solution with a non vanishing order parameter
of the 1RSB type
might lead to a thermodynamically more favorable phase.

The 1RSB solution corresponds to group the $n$ replicas into $n/m$
clusters of $m$ replicas. Any two replicas $\alpha\not=\beta$ within
the same cluster have overlap $q_1$, whereas replicas in different
clusters have overlap $q_0< q_1$. As a consequence the $n\times n$
${\bm q}$ matrix breaks down into $(n/m)\times (n/m)$ blocks of
dimension $m\times m$. If the element $q_{\alpha\beta}$ with
$\alpha\not=\beta$ belongs to one of the diagonal block then
$q_{\alpha\beta}=q_1$, otherwise $q_{\alpha\beta}=q_0$. The overlap
matrix for the 1RSB {\it ansatz} can be conveniently written as:
\begin{equation}
 q_{\alpha\beta} = (1-q_1)\,\delta_{\alpha\beta} + 
                   (q_1 - q_0)\,\epsilon_{\alpha\beta} + q_0
\end{equation}
where the matrix ${\bm \epsilon}$ is defined as
\begin{equation}
 \epsilon_{\alpha\beta} = \left\{\begin{array}{ll}
         1 & \mbox{if $\alpha$ and $\beta$ are in a diagonal block}\\
         0 & \mbox{otherwise} \\
        \end{array}
        \right.
\end{equation}
By plugging this form of $q_{\alpha\beta}$ into the Eq. (\ref{eq:free2}) 
one obtains:
\begin{eqnarray}
\label{eq:g1rsb}
2\lim_{n\to 0}\frac{1}{n}G[{\bm q}] &=& g(1) - g(q_1) + m\, [g(q_1) - g(q_0)] 
            \nonumber \\
      &\phantom{=}& 
           +\frac{q_0}{\chi(q_0)} 
           + \frac{1}{m}\,\ln \chi(q_0)
      \nonumber \\
      &\phantom{=}& 
           +\frac{m-1}{m}\,\ln\chi(q_1).
\end{eqnarray}
where, for later convenience, we have defined\cite{note9}
\begin{eqnarray}
\chi(q_1) &=& 1 - q_1 \\
\chi(q_0) &=& 1 - q_1 + m\,(q_1 - q_0).
\end{eqnarray}

The saddle point equations for $q_0$ and $q_1$ in the limit $n\to 0$, 
obtained either from (\ref{eq:spab}) or directly from stationarity of 
(\ref{eq:g1rsb}) with respect to variations of $q_0$ and $q_1$, read 
\begin{eqnarray}
\label{eq:q01rsb}
\Lambda(q_0) &-& \frac{q_0}{\chi(q_0)^2} = 0 \\
\label{eq:q11rsb}
\Lambda(q_1) &-& \Lambda(q_0) - 
  \frac{q_1 - q_0}{\chi(q_1)\, \chi(q_0)} = 0
\end{eqnarray}
The solution of these equations depends on the value of $m$ that,
 in the limit $n\to 0$, is restricted to the interval $0\leq m\leq 1$.
In principle any value of $m$ which leads to a stable 1RSB solution can be 
chosen. However in the spirit of the saddle point calculation performed to
evaluate the free energy we choose for any value of $\mu_p$'s 
the value of $m$ which minimize the functional $G[{\bm q}]$.\cite{note4} 
This leads to the additional equation
\begin{eqnarray}
\label{eq:m1rsb}
g(q_1) - g(q_0) &+& \left[
                 \frac{1}{m\,\chi(q_0)} - \frac{q_0}{\chi(q_0)^2}
                    \right]\, (q_1 - q_0)
    \nonumber\\
   &+& \frac{1}{m^2}\ln\left[\frac{\chi(q_1)}{\chi(q_0)}\right] = 0.
\end{eqnarray}

The stability analysis of the 1RSB saddle shows that 
in the limit $n\to 0$ the 1RSB solution is stable  as 
long as the 1RSB eigenvalues\cite{note7}
\begin{eqnarray}
\label{eq:rep1rsb-1}
\Lambda_1^{(1)} &=& -\Sigma(q_1) + \frac{1}{\chi(q_1)^2} \\
\label{eq:rep1rsb}
\Lambda_0^{(3)} &=& -\Sigma(q_0) + \frac{1}{\chi(q_0)^2}
\end{eqnarray}
are both positive.

The saddle point equation (\ref{eq:q01rsb}) admits always the solution
$q_0=0$. It may also have solutions with $0<q_0<1$, however
by using arguments similar to those that lead to the 
inequality (\ref{eq:qunsta}) for the RS solution, 
one can show that in absence of external field any 1RSB solution
with $q_0>0$ is unstable since it has a negative $\Lambda_0^{(3)}$.\cite{note5}

The 1RSB saddle point equations for $q_1$ and $m$  
can be solved for any $p$ using the same procedure used
for the spherical $p$-spin spin glass model. The first step is to obtain
$g(q_1)$ from equation (\ref{eq:m1rsb}) [with $q_0=0$] and divide it by 
$q_1\,\Lambda(q_1)$. Then using the saddle point equation
(\ref{eq:q11rsb}) [with $q_0=0$]
to express $\Lambda(q_1)$ one ends up with the equation
\begin{equation}
\label{eq:z1rsb}
2\ \frac{g(q_1)}{q_1\,\Lambda(q_1)} = z(y)
\end{equation}
where
\begin{equation}
\label{eq:CS-z}
z(y) = -2y \frac{1-y + \ln y}{(1-y)^2}
\end{equation}
is the auxiliary $z$-function introduced 
by Crisanti and Sommers\cite{CriSom92} (CS)
for the solution of the spherical $p$-spin spin glass model, and
\begin{equation}
\label{eq:y1rsb}
y \equiv \frac{\chi(q_1)}{\chi(q_0)}=
 \frac{1-q_1}{1-q_1 + m\,q_1}, \qquad 0\leq y \leq 1.
\end{equation}
By using $y$ and $m$ as free parameters
eqs. (\ref{eq:q11rsb}) [with $q_0=0$] and (\ref{eq:z1rsb})
can be solved for $(\mu_p,\mu_2)$. A straightforward algebra leads to:
\begin{eqnarray}
\label{eq:mup1rsb}
\mu_p &=& \frac{p}{(p-2)}\, \frac{[1 - z(y)]}{q_1^{p-2}(1-q_1)(1-q_1 + m\,q_1)}
\nonumber\\
      &=& \frac{p}{(p-2)}\,\frac{(1-y+m\,y)^p}{m^2 y(1-y)^{p-2}}\,[1-z(y)],
\\
\label{eq:mu21rsb}
\mu_2 &=& \frac{1}{(p-2)}\, \frac{[p\,z(y) -2]}{(1-q_1)(1-q_1 + m\,q_1)}
\nonumber\\
      &=& \frac{(1-y+m\,y)^2}{m^2 y}\,\frac{[p\,z(y) -2]}{(p-2)}.
\end{eqnarray}
By fixing the value of $m$ in the interval $[0,1]$ and varying $y$ 
these equations represent the parametric equations of the so called 
$m$-lines in the $(\mu_p,\mu_2)$ plane. By definition $y$ can take any value
between $0$ and $1$ included, however, from 
Eq. (\ref{eq:mu21rsb}) we see that since $z(0)=0$, $\mu_2$ becomes negative 
for $y$ sufficiently close to $0$.
Setting $\mu_2=0$ from Eq. (\ref{eq:mu21rsb}) one gets
\begin{equation} 
 p\,z(y) - 2 = 0
\end{equation}
which gives the minimum value $y_{\rm min}$ of $y$. 
The CS $z$-function (\ref{eq:CS-z}) is a monotonous increasing function of $y$ 
varying in the range $0\leq z\leq 1$, as a consequence, $\mu_p$ is always
non-negative. 

A second condition on $y$ comes form the stability analysis of the 1RSB 
solution. 
A simple inspection shows that $\Lambda_1^{(1)}>\Lambda_0^{(3)}$ 
so that the condition which marks the limit of the stability of the 
1RSB solutions is $\Lambda_0^{(3)} = 0$, i.e., 
\begin{equation}
\label{eq:uns1rsb}
\mu_2 = \frac{1}{(1-q_1+m\,q_1)^2}
\end{equation}
Using now Eq. (\ref{eq:mu21rsb}) one gets the equation
\begin{equation}
\label{eq:ymax1rsb} 
        p\,z(y) - 2 - (p-2)\, y= 0
\end{equation}
whose solution gives $y_{\rm max}$, the maximum  value of $y$ 
for the 1RSB solution.

Both boundary values $y_{\rm min}$ and $y_{\rm max}$ are functions of
$p$ only. For example, for $p=3$, we have
\begin{equation}
y_{\rm min} = 0.354993...,\quad  y_{\rm max} = 1 
\end{equation}
while for $p=4$
\begin{equation}
y_{\rm min} = 0.195478...,\quad  y_{\rm max} =  0.389571...
\end{equation}
The fact that for $p=3$ the maximum is $y_{\rm max} = 1$ 
makes the $2+3$ model different from any other $2+p$ model with $p>3$,
as we shall see in a while.

From the stability condition (\ref{eq:uns1rsb}) it follows that if the
quadratic term in the Hamiltonian were
missing, as for the already mentioned  $3+p$ model, then
the 1RSB solution would be stable everywhere. In the Appendices \ref{app:s+p}
and \ref{app:stabil} we shall show that indeed in this case the 1RSB solution
is the only possible non-trivial solution, besides the RS solution.


\subsubsection{The Transition Lines between the Paramagnet and the 1RSB-Glass phase}
To find the transition lines which bound the 1RSB phase we start by
noting that the $m$-lines do not cross and that the
value of $\mu_2$ for which the 1RSB becomes unstable increases as $m$ 
decreases, see Eq. (\ref{eq:uns1rsb}). Moreover, all $m$-lines start
from $\mu_2=0$. 
As a consequence the first $m$-line one encounters in moving from the
RS phase at fixed $\mu_2<1$ and increasing
$\mu_p$ is the $m$-line with $m=1$.
This line, which marks the transition between the RS (paramagnetic) phase
and the 1RSB (glass) phase, 
starts on the $\mu_p$ axis at the point $\mu_p=\mu_p(y_{\rm min},m=1)$ 
and goes up to the point $\mu_p=\mu_p(y_{\rm max},m=1)$ and $\mu_2=1 $,
as can be easily seen from Eq. (\ref{eq:uns1rsb}) evaluated for $m=1$.

The transition between the RS ($q=0$) and the 1RSB 
($q_1\not=0$, $q_0=0$) phases is not due to an instability
but occurs because the 1RSB solution leads to a thermodynamically 
more favorable 
state.  Since we are dealing with the replica trick, this means that 
the 1RSB solution yields a value of the free energy functional 
(\ref{eq:free1}) {\em larger} than the RS solution.\cite{note4} 

This mechanism resembles that of ordinary first order
transitions, and indeed the order parameter $q_1$ jumps
discontinuously from zero to a finite value, and vice-versa, at the
transition. However, the free energy remains continuous across the
transition --at $m=1$ the free energies of the two solutions are equal--
and no discontinuity occurs in its first derivatives.

The 1RSB solution becomes unstable when $\Lambda_0^{(3)}=0$. This leads 
to a second transition line whose parametric equation in the 
$(\mu_p,\mu_2)$ plane is 
obtained by setting $y=y_{\rm max}$ into eqs.
(\ref{eq:mup1rsb})-(\ref{eq:mu21rsb}) and varying $m$ from $1$,
 to $0$.
For $m\to 0$ the values of  both $\mu_p$ and $\mu_2$ diverge but
\begin{equation}
\lim_{m\to 0} \frac{\mu_2}{\mu_p} =
                \frac{(1-y_{\rm max})^{p-2}\, [2 - (p-2)y_{\rm max}]}
                     {(p-2)\,y_{\rm max} - p}
\end{equation}
and hence the 1RSB phase does not cover the full ``low temperature'' phase
of the model.

\begin{figure}
\includegraphics[scale=1.0]{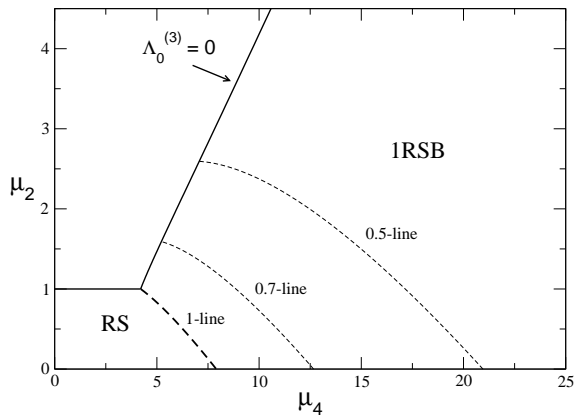}
\caption{The RS and 1RSB phases for the $2+4$ model in the $(\mu_4,\mu_2)$ 
         plane. The thick lines are the transition lines
	 between different phases.
        }
\label{fig:p41rsb}
\end{figure}

In Figure \ref{fig:p41rsb} the transition lines found so far 
are shown together with the $m$-lines with $m=0.7, 0.5$. 
In the figure $p=4$, but any $p>3$ leads to a qualitatively similar
scenario.

\begin{figure}
\includegraphics[scale=1.0]{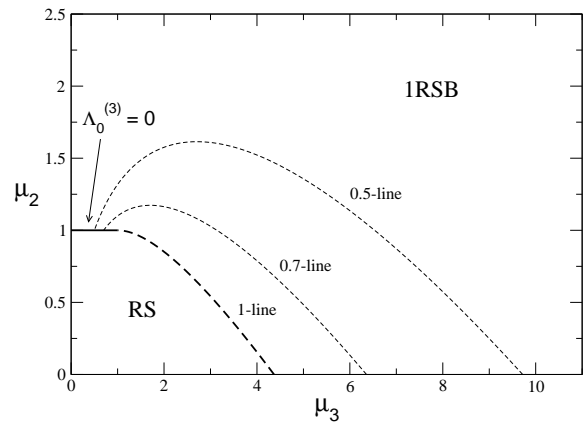}
\caption{The phase diagram of the $2+3$ model in the $(\mu_3,\mu_2)$
         plane. The thick lines are the transition lines between the 
         RS and 1RSB phases. In particular the thick dashed line, the
         $m$-line with $m=1$, is the discontinuous transition while the
         horizontal thick full line is the continuous transition.
        }
\label{fig:p31rsb}
\end{figure}

The case $p=3$ is special because inserting $y=y_{\rm max}=1$ into 
eqs. (\ref{eq:mup1rsb})-(\ref{eq:mu21rsb}) one ends up
with 
\begin{equation}
\label{eq:crit23}
\left\{\begin{array}{lll}
\mu_3 &=& m \\
\mu_2 &=& 1\\
\end{array}\right. \qquad 0\leq m \leq 1
\end{equation}
Along this line $q_1=0$, see e.g. Eq. (\ref{eq:y1rsb}), and the 1RSB solution
reduces to the RS solution. We have seen that the RS solution 
becomes unstable for $\mu_2=1$ thus the critical line (\ref{eq:crit23})
marks the transition between the RS and the 1RSB phases. 
The transition is continuous in both the 
free energy and the order parameter $q_1$. 
The transition lines for the $2+3$ model are shown in Figure \ref{fig:p31rsb}.
In conclusion
the $2+3$ model presents only one ``low temperature'' phase of 1RSB type and,
in this respect, is equivalent to the spherical $p$-spin spin glass model
in a field.\cite{CriSom92} 


\subsection{The One-Full Replica Symmetry Broken Solutions (1-FRSB)}
From Figure \ref{fig:p41rsb} one clearly sees that for $p>3$ the RS and 1RSB 
solutions do not cover the whole phase space of the $2+p$ model. 
In the region where both the RS and 1RSB solutions 
are unstable the organization of pure 
states has a more complex structure which cannot be described 
by a simple 1RSB {\it ansatz} which groups them into $n/m$ equivalent 
clusters. 
Therefore, to describe this region one must allow for clusters 
of different type.
To this end each one of the $n/m$ clusters is divided it into 
$m/m_1$ sub-cluster of size $m_1$. If the procedure is repeated $R$ 
times, dividing at each step the smallest clusters into yet smaller clusters,
one has the $R$-RSB {\it ansatz} in which the replica symmetry
is broken $R$-times. The overlap $q_{\alpha\beta}$ between two replicas
depends on the number of divisions  separating the offspring clusters to which
the replicas 
$\alpha$ and $\beta$ belong  from the common ancestor cluster.

A simple way proposed by Parisi\cite{Parisi80} to parametrize the overlap 
matrix 
$q_{\alpha\beta}$ for $R$ steps in the replica symmetry breaking 
consists in dividing ${\bm q}$ into successive boxes of decreasing size $p_r$, 
with $p_0 = n$ and $p_{R+1}=1$, and assigning the elements 
$q_{\alpha\beta}$ of the matrix ${\bm q}$ so that 
\begin{equation}
q_{\alpha\beta} \equiv q_{\alpha\cap\beta=r} = q_r, \qquad r = 0,\cdots, R+1
\end{equation}
with $1 = q_{R+1} > q_R >\cdots q_1 > q_0$. The notation 
$\alpha\cap\beta=r$ means that $\alpha$ and $\beta$ belong to the 
same box of size $p_r$ but to two {\it distinct} boxes of size $p_{r+1} < p_r$.

Inserting this form of $q_{\alpha\beta}$ into Eq. (\ref{eq:free2}) 
one gets with standard manipulations
\begin{eqnarray}
\frac{2}{n}\, G[{\bm q}] &=& g(1) 
                         + \sum_{r=0}^{R}\, (p_r - p_{r+1})\, g(q_r) 
\nonumber\\
&\phantom{=}&
                         + \ln(1 - q_R) 
                         + \sum_{t=0}^{R}\, \frac{1}{p_r}\,
                            \ln\frac{\hat{q}_r}{\hat{q}_{r+1}}
\label{eq:free-rsb}
\end{eqnarray}
where $\hat{q}_r$ is the Replica Fourier Transform
of $q_{\alpha\beta}$,\cite{Cris,DeDom}
\begin{equation}
\hat{q}_r = \sum_{s=r}^{R+1}\, p_s\, (q_s - q_{s-1})
\end{equation}

The number $R$ is arbitrary. Setting $R=0$ or $R=1$ one recovers 
respectively the RS and the 1RSB expressions, the latter with $m = p_1$, while
for $R\to\infty$ one gets the $\infty$-RSB solution or 
Full Replica Symmetry Broken (FRSB) solution.
In this limit the differences $p_{r+1} - p_{r}$ become
infinitesimal and the set of overlaps $\{q_0,...,q_{R}\}$
is replaced in the limit $n\to 0$ by a non-decreasing continuous 
function $q(x)$ defined on the interval $x\in[0,1]$.

The free energy functional (\ref{eq:free-rsb}) for the Parisi 
$R$-RSB  {\it ansatz} can be conveniently expressed by using the function
\begin{equation}
\label{eq:xqr}
x(q) = p_0 + \sum_{r=0}^{R} (p_{r+1} - p_r)\, \theta(q - q_r)
\end{equation}
which equals the fraction of pair of replicas with overlap $q_{\alpha\beta}$ 
less or equal to $q$. 
With this definition, and replacing the sums by integrals, one obtains, after
a little of algebra
\begin{eqnarray}
\frac{2}{n}\, G[{\bm q}] &=& 
            \int_{0}^{1} dq~ x(q)\, \Lambda(q) 
\nonumber\\
&\phantom{=}&
            + \int_{0}^{q_R} \frac{dq}{\int_{q}^{1} dq'\, x(q')}
            + \ln\left(1 - q_R\right)
\label{eq:rpl-f}
\end{eqnarray}
This expression is valid for any $R$, and hence also for
the FRSB solution.
In the limit $R\to\infty$, $q_r$ becomes continuous and
we can define $q(x)$ as the inverse of $x(q)$. It can be shown
that $dx(q)/dq$ gives 
the probability density of overlaps.\cite{Parisi79,Parisi80}

It is easy to verify that taking for $q(x)=0$ or $q(x) = q_1\,\theta(x-m)$ 
the above functional reduces to those found with the RS and 1RSB 
{\it ansatzs}, respectively.\cite{note1}

The FRSB solution with a {\em continuous} $q(x)$ was introduced to 
describe the spin-glass phase of the SK model,\cite{Parisi80} 
and since then it has been found 
in many other related models.
A continuous order parameter function $q(x)$ is, however,
not general enough to describe the state of the
$2+p$ model with $p>3$ in the whole parameter space. 
From the stability analysis of the 1RSB solution we see indeed that the 
instability occurs because the eigenvalue $\Lambda_0^{(3)}$ vanishes. 
This eigenvalue is associated with fluctuations that involve the overlaps
of one cluster as a whole with the other clusters as a whole.\cite{CriSom92}
Roughly speaking these fluctuations are similar to fluctuations in the
RS phase with single replicas replaced by 
the clusters of $m$ replicas considered as single entities.
As a consequence we expect that, as it happens for the fluctuations in the 
RS phase, a non zero overlap $q_0$ between clusters would stabilize the
fluctuations. The solution however cannot be of 1RSB type since we have 
seen that any 1RSB solution with $q_0\not=0$ is unstable.

\begin{figure}
\includegraphics[scale=1.0]{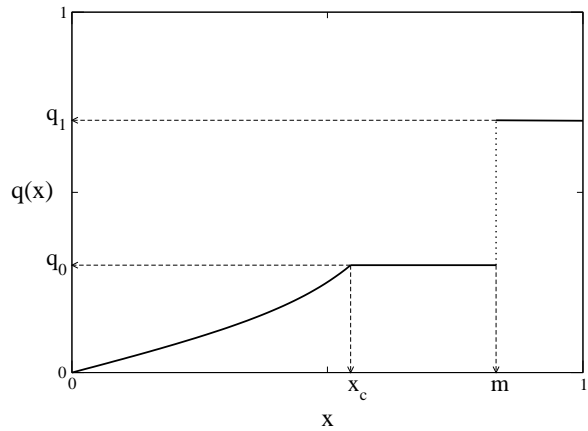}
\caption{Schematic form of the order parameter function $q(x)$ in the
1-FRSB phase.
}
\label{fig:qx-1frsb}
\end{figure}

Based on the analogy with the instability of RS solution 
with clusters playing the role
of single replicas, it turns out that the correct 
{\it ansatz} for the $2+p$ model with $p>3$ is a mixture of 1RSB and FRSB, 
which we have called 1-FRSB solution,\cite{CriLeu04} 
described by a {\em discontinuous} order
parameter function in the interval $[0,1]$ of the form
\begin{equation}
\label{eq:1frsb-qx}
q(x) \Rightarrow \left\{ \begin{array}{ll}
  q_1   & \mbox{for $x>m$} \\
  q(x)  & \mbox{for $x<m$} 
  \end{array}
\right.
\end{equation}
where $q(x)$ is a non-decreasing continuous function in the semi-open 
interval 
$x\in[0,m)$, with $\lim_{x\to m^-}q(x) = q_0 < q_1$
and $q(0) = 0$,\cite{note5} 
see Fig. \ref{fig:qx-1frsb}.
For $q_0=q_1$ one recovers the FRSB solution.\cite{note8}
In appendix \ref{app:stabil} we show that the 1-FRSB solution is the
only other possible non-trivial solution, beside the 1RSB 
(modeling a mean-field
glass) and the FRSB (modeling a spin-glass) ones, for the $2+p$ model
with $p>3$. It is interesting to note that this solution also follows
by solving numerically in the whole interval $[0,1]$ 
the Parisi equations derived from the stationarity 
of the functional
(\ref{eq:rpl-f}) with respect to order parameter function $q(x)$,
see Appendices \ref{app:SomDup}, \ref{app:SomDup-sol}. The
partial differential equation is solved numerically
by means of a pseudo spectral technique, see
e.g. Ref. [\onlinecite{CLP02}], without fixing {\em a priori} 
any special {\em ansatz} for $q(x)$.

The free energy functional for the 1-FRSB {\it ansatz} can
be obtained either by inserting the explicit form (\ref{eq:1frsb-qx})
of $q(x)$ into Eq.  (\ref{eq:rpl-f}), or by taking in Eq. (\ref{eq:free-rsb})
$p_R=m$, $q_R=q_1$, $q_{R-1}=q_0$ and $p_R - p_{R-1}$ finite as 
$R\to\infty$. In both cases one ends up with
\begin{eqnarray}
\frac{2}{n}\, G[{\bm q}] &=& g(1) - g(q_1) + m\,[g(q_1) - g(q_0)]
\nonumber\\
&\phantom{=}&
            + \int_{0}^{q_0} dq\, x(q)\, \Lambda(q)
            + \int_{0}^{q_0} \frac{dq}{\chi(q)}
\nonumber\\
&\phantom{=}&
            + \frac{m-1}{m}\ln\left(1 - q_1\right)
	    + \frac{1}{m}\ln\chi(q_0)
\label{eq:1frsb-f}
\end{eqnarray}
where
\begin{equation}
\label{eq:chi}
\chi(q) = 1 - q_1 + m\,(q_1 - q_0) + \int_{q}^{q_0}\,dq'\, x(q')
\end{equation}

Stationarity of the free energy functional $\Phi$ with respect to
$q(x)$ and $q_1$ leads to the 1-FRSB saddle point equations:
\begin{equation}
\label{eq:qx1frsb}
\Lambda(q) = \int_{0}^{q}\, \frac{dq}{\chi^2(q)}, \qquad 0\leq q\leq q_0
\end{equation}
and
\begin{equation}
\label{eq:q11frsb}
\Lambda(q_1) - \Lambda(q_0) = \frac{q_1 - q_0}{\chi(q_1)\,\chi(q_0)}.
\end{equation}
Finally maximization with respect to $m$ leads to the additional
equation
\begin{eqnarray}
\label{eq:m1frsb}
 g(q_1) - g(q_0) &=& 
   - \left[\frac{1}{m\,\chi(q_0)} - \Lambda(q_0)\right] (q_1 - q_0)
\nonumber\\
&\phantom{=}& - \frac{1}{m^2}\,\ln\left[\frac{\chi(q_1)}{\chi(q_0)}\right]
\end{eqnarray}

The 1-FRSB saddle point equations (\ref{eq:q11frsb}) and (\ref{eq:m1frsb})
are formally equal to the 1RSB saddle point equations 
(\ref{eq:q11rsb}) and (\ref{eq:m1rsb}) and hence can be solved 
for any $p$ with the help of the CS $z$-function.  Indeed 
by using eqs. (\ref{eq:q11frsb}) and (\ref{eq:m1frsb}) it is easy to 
verify that 
\begin{equation}
\label{eq:z1frsb}
2\, \frac{g(q_1) - g(q_0) - (q_1 - q_0)\, \Lambda(q_0)}
       {(q_1 - q_0)\, [\Lambda(q_1) - \Lambda(q_0)]} = z(y)
\end{equation}
where $z(y)$ is given by Eq. (\ref{eq:CS-z}) and  $y$ is defined as
\begin{equation} 
\label{eq:auxy}
y = \frac{\chi(q_1)}{\chi(q_0)} \equiv \frac{1-q_1}{1 - q_1 + m\,(q_1 - q_0)]},
\quad 0\leq y\leq 1.
\end{equation}

The saddle point equation (\ref{eq:qx1frsb}) is not easy to use as it stands.
Differentiating both sides with respect to $q$ to eliminate the integral 
one gets the more manageable form:
\begin{equation}
\label{eq:rep1frsb}
\Sigma(q) = \frac{1}{\chi(q)^2}
\end{equation}

Equations (\ref{eq:q11frsb}), (\ref{eq:z1frsb}) and (\ref{eq:rep1frsb})
evaluated for $q=q_0$ can be solved for $(\mu_p,\mu_2)$ 
as function of $m$ and $t = q_0 / q_1$.
After a straightforward algebra one ends up with
\begin{eqnarray}
\label{eq:mup1frsb}
\mu_p &=& \frac{[1-y+my(1-t)]^p}
       {m^2 y (1-y)^{p-3} (1-t)} 
\nonumber\\
&\phantom{=}& \times
     \frac{1}{\left[ 1 - (p-1)t^{p-2} + (p-2) t^{p-1}\right]},
\\ 
\label{eq:mu21frsb}
\mu_2 &=& \frac{[1-y+my(1-t)]^2}
       {m^2 y (1-t)^2 }
\nonumber\\
&\phantom{=}& \times
   \frac{\left[y(1-t^{p-1}) (p-1)(1-t)t^{p-2}\right]}
       {\left[ 1 - (p-1)t^{p-2} + (p-2) t^{p-1}\right]},
\end{eqnarray}         
where for any $0\leq t\leq 1$, $y$ is solution of the equation
\begin{eqnarray}
\label{eq:yt1frsb}
&&p(1-t)\bigl[ 1 - (p-1)t^{p-2} + (p-2) t^{p-1}\bigr]\, z(y)  \nonumber \\
&-&
   \bigl[p - 2 - p t + p t^{p-1} - (p-2) t^p\bigr]\,y  - 2
+ p(p-1)t^{p-2} \nonumber \\
&-& 2 p (p-2) t^{p-1} + (p-1)(p-2)t^p=0
\end{eqnarray}

Equations (\ref{eq:mup1frsb})-(\ref{eq:yt1frsb}) are the parametric equations
of the 1-FRSB $m$-lines which are drawn in the $(\mu_p,\mu_2)$ plane 
by fixing the value of $m$ in the interval $[0,1]$ and varying $t$ from 
$0$ to $1$.
The 1-FRSB $m$-line begins for $t=0$ at the boundary with the 1RSB phase 
and ends for $t=1$ where the 1-FRSB solution goes over to the FRSB solution.
The values of $y$ for this limiting cases are:
\begin{eqnarray}
\label{eq:ymin-1frsb}
t = 0\ &\Rightarrow&\ p z(y) - 2 - (p -2)\, y = 0 \\
\label{eq:ymax-1frsb}
t = 1\ &\Rightarrow&\ y = 1
\end{eqnarray}
By comparing Eq. (\ref{eq:ymin-1frsb}) with Eq. (\ref{eq:ymax1rsb})
one recognize that the value of $y$ for $t=0$ 
is equal to the maximum allowable value of $y$ for the 1RSB $m-$lines.
As a consequence 1-FRSB $m$-lines and the 1RSB $m$-lines with the 
{\em same} $m$
match continuously at the transition point between the two solutions.

By evaluating Eq. (\ref{eq:rep1frsb}) for $q=q_0$ it is easy to see that the
eigenvalue
$\Lambda_0^{(3)}$ [Eq. (\ref{eq:rep1rsb})] is identically zero in the
whole 1-FRSB (and FRSB) phase, in agreement with the marginal stability
of FRSB solutions.\cite{DeDomKon} The eigenvalue $\Lambda_1^{(1)}$ 
[Eq. (\ref{eq:rep1rsb-1})] remains positive in the whole 
1-FRSB phase and vanishes for $t=1$ where the 1-FRSB solutions disappears 
in favor of the FRSB solution. 

\begin{figure}
\includegraphics[scale=1.0]{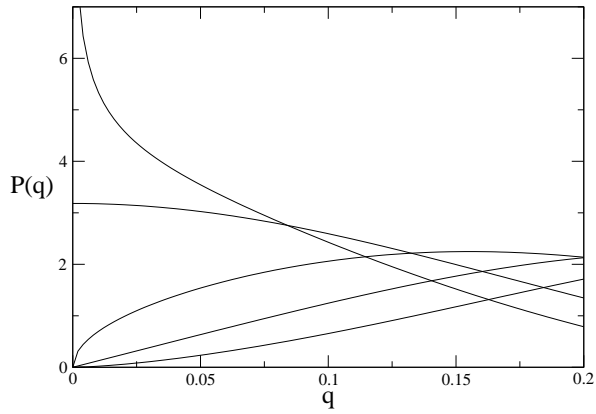}
\caption{
  The probability distributions of the overlap are plotted for different
values of $p$.  The values for the other parameters are $\mu_2 = 2,
\mu_p = 3$ and $q_0 =0.2$. The qualitative picture  for small $q$'s
only depends,
anyhow, on the value of $p$.
Referring to the vertical ax, from top to bottom
the $P_p(q)$ for $p=3.5,4,4.5,5$ and $5.5$ are reproduced.
At  $p = 3.5$ (top curve) the $P(q)$ diverges at $q=0$,
for $p=4$ it goes to a finite value. When $p>4$ it tends to zero as $q\to 0$.}
\label{fig:pq}
\end{figure}

The continuous part $q(x)$ of the order parameter function can be obtained
from Eq. (\ref{eq:rep1frsb}). Indeed from this equation it follows that
\begin{eqnarray}
1 - q_1 &+& m\,(q_1 - q_0) + \int_{q}^{q_0}\, dq'\, x(q') = \nonumber\\
&\phantom{+}&\frac{1}{\sqrt{\mu_2 + \mu_p (p-1) q^{p-2}}}
\end{eqnarray}
which differentiated with respect to $q$ leads to the sought solution
\begin{equation}
\label{eq:xqfrsb}
 x(q) = \frac{\mu_p}{2} 
\frac{(p-1)\,(p-2)\, q^{p-3}}{\bigl[\mu_2 + \mu_p\,(p-1)\,q^{p-2}\bigr]^{3/2}},
\quad 0\leq q\leq q_0.
\end{equation}
We note that as $q\to 0$ the probability density of the overlaps $dx(q)/dq$ 
goes as $q^{p-4}$ so that it
diverges for $3<p<4$, is finite for $p=4$ and vanishes for $p>4$ (see
Fig. \ref{fig:pq}).

Unlike the SK case\cite{CriRiz02} the function $q(x)$ for the
$2+p$ model with $p>3$ is not 
a linear function of $x$ for $x\ll 1$. From the solution 
(\ref{eq:xqfrsb}) it is easy to see that
\begin{equation}
q(x) \sim x^{1/(p-3)}, \qquad x\to 0
\end{equation}
so that only for $p=4$ one recovers a linear behavior.\cite{CriLeu04} 
As a consequence $dq(x)/dx$ vanishes for
$x\to 0$ for $3<p\leq 4$ and diverges for $p>4$.

The function $q(x)$ for a generic $p$ can be obtained by 
expanding the r.h.s of 
Eq. (\ref{eq:xqfrsb}) in powers of $q^{p-2}$ and then inverting the series.
As an example we give the first few terms for the case 
$p=4$:\cite{note6}
\begin{eqnarray}
q(x) &=&   \frac{\mu_2^{3/2}}{3\mu_4}\, x
       + \frac{\mu_2^{7/2}}{6\mu_4^4}\,x^3
       + \frac{13 \mu_2^{11/2}}{72\mu_4^3}\,x^5 
\nonumber \\
&\phantom{=}&
       + \frac{323 \mu_2^{15/2}}{1296\mu_4^4}\,x^7
       + \frac{4025 \mu_2^{19/2}}{10368\mu_4^5}\,x^9 
       + O(x^{10})
\end{eqnarray}
and $p=5$:
\begin{eqnarray}
q(x) &=&   \sqrt{\frac{\mu_2^{3/2}}{\mu_5}}\, x^{1/2}
         + \frac{\mu_2^{2}}{126\mu_5}\,x^2 \nonumber \\
&\phantom{=}& 
       + \frac{17}{144} \sqrt{\frac{\mu_2^{13/2}}{6\mu_5^3}}\,x^{7/2} 
       + O(x^{9/2})
\end{eqnarray}

\begin{figure}
\includegraphics[scale=1.0]{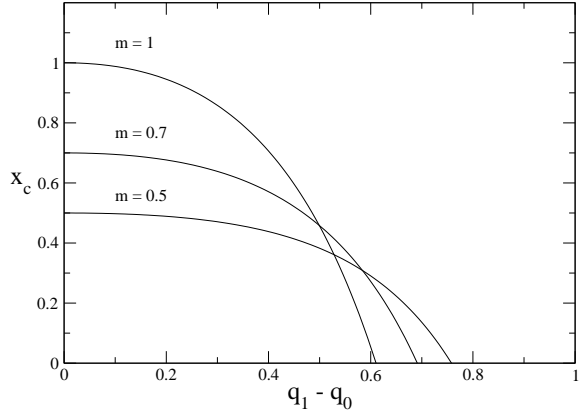}
\caption{$x_c=x(q_0)$ versus the $q_1 - q_0$ 
         in the 1-FRSB phase for $p=4$ and $m=0.5, 0.7, 1$.
        }
\label{fig:xc-p41frsb}
\end{figure}

\begin{figure}
\includegraphics[scale=1.0]{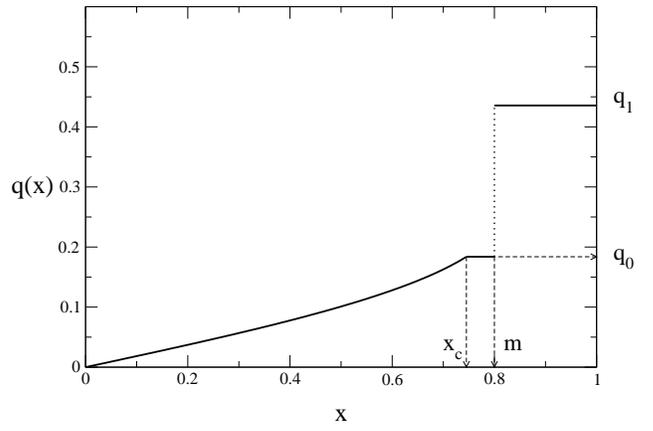}
\caption{The order parameter function $q(x)$ in the 1-FRSB phase of the
         $2+4$ model. In the figure
         $m=0.8$, $\mu_4 = 3$ and $\mu_2 = 1.4$.
}
\label{fig:qx-2+4-1.4-3}
\end{figure}

The continuous part of the order parameter functions ends for $q=q_0$ at the
point $x_c = x(q_0)$. In the 1-FRSB phase $x_c$ 
is always smaller than $m$ and becomes equal to it at the boundary line
with the FRSB phase. In Figure \ref{fig:xc-p41frsb} we show the value of $x_c$ 
as function of the difference $q_1-q_0$ for a fixed value of $m$.
For values of $x$ between 
$x_c$ and $m$ the order parameter function $q(x)$ remains constant and equal
to $q_0$, and then jumps to $q_1$ as $x$ goes through $m$,  
see Figure \ref{fig:qx-2+4-1.4-3}.

\subsubsection{The Transition Lines among the Amorphous Phases (1RSB, 1-FRSB and FRSB)}
We have seen that the 1-FRSB $m$-lines are the continuations
into the 1-FRSB phase of the 1RSB $m$-lines. As a consequence, as the 1RSB 
$m$-line with $m=1$ marks the transition between the 1RSB phase and RS phase, 
so the 1-FRSB $m$-line with $m=1$ marks the transition between the 
1-FRSB phase and the FRSB phase. 
The transition is discontinuous in the order parameter since 
$q_1-q_0$ does not vanish at the transition, but the discontinuity 
appears for $m=1$ and the free energy and its derivatives  
remain continuous across the transition. 

The 1-FRSB $m$-line with $m=1$ ends at the critical point ($t=1$)
\begin{eqnarray}
\label{eq:mupcrit}
\mu_p^{*} &=& \frac{2}{27}\, \frac{p^p}{(p-1)(p-2)(p-3)^{p-3}} \\
\label{eq:mu2crit}
\mu_2^{*} &=& \frac{1}{(p-2)}\, \left(\frac{p}{3}\right)^3
\end{eqnarray}
where
\begin{equation}
 q_0 = q_1 = \frac{p-3}{p}.
\end{equation}
For $\mu_2 > \mu_2^*$ the transition between the 1-FRSB phase
and the FRSB takes place continuously in the order parameter function
with $q_1-q_0\to 0$ and $x_c\to m$ at the transition. 
The continuous transition 
between the 1-FRSB and the FRSB phases occurs on the line of end points
of the 1-FRSB $m$-lines. 
Inserting $t=1$ into eqs. (\ref{eq:mup1frsb})-(\ref{eq:mu21frsb})
one easily gets the parametric equations of the critical line:
\begin{eqnarray}
\mu_p &=& \frac{2\,(p-3 + 3\,m)^p}{27\,m^2\,(p-1)\,(p-2)\,(p-3)^{p-3}} \\
\mu_2 &=& \frac{p\,(p-3 + 3\, m)^2}{27\,m^2(p-2)}
\end{eqnarray}
where $0\leq m\leq 1$. Along this line $x_c = m$, $\Lambda_1^{(1)} =0$ and
\begin{equation}
 q_0 = q_1 = \frac{p-3}{p-3 + 3\,m}, \qquad 0\leq m\leq 1
\end{equation}

Finally the 1-FRSB phase is bounded by the transition 
line with the 1RSB phase. Indeed by setting $t=0$ into eqs. 
(\ref{eq:mup1frsb})-(\ref{eq:mu21frsb}) one recovers the parametric equations
of the 1RSB instability line: $\Lambda_0^{(3)}=0$ and $q_0=0$. 
The transition is continuous in both free energy
and order parameter function since $q_0\to 0$ continuously as the transition
line is approached from the 1-FRSB side.

\begin{figure}
\includegraphics[scale=1.0]{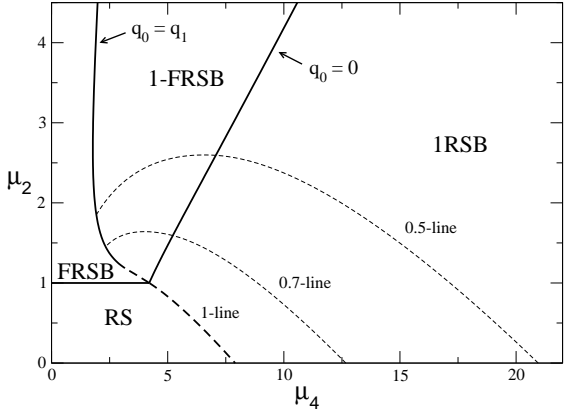}
\caption{The static phase diagram of the $2+4$ model in the $(\mu_4,\mu_2)$ 
         plane.}
\label{fig:p41frsb}
\end{figure}

All the transition lines, together with the $m$-lines with $m=0.7$ and 
$m=0.5$, and the phases of the $2+p$ model with $p>3$ are shown in 
Figure \ref{fig:p41frsb}. In the figure $p=4$, but the phase diagram
does not change qualitatively with the value of $p$, provided that it remains
larger than $3$.

\begin{figure}
\includegraphics[scale=1.0]{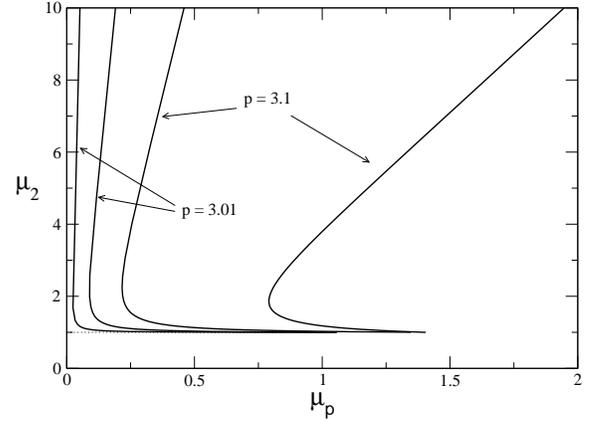}
\caption{Borders lines of the 1-FRSB and FRSB phases for $p=3.01$ and $p=3.1$.
        }
\label{fig:frsb-p}
\end{figure}

In the limit $p\to 3$ the 1-FRSB and FRSB phases shrink to zero
while the transition lines separating the two phases collapse smoothly onto 
the vertical line $(0,\mu_2)$ with $\mu_2\geq 1$ and the horizontal line
$(1,\mu_p)$ with $0\leq \mu_p  \leq 1$ where $q_0=q_1=0$, see 
Figure \ref{fig:frsb-p}.  One then 
smoothly recovers the phase diagram of the $2+3$ model, 
Figure \ref{fig:p31rsb}.

From Figure \ref{fig:frsb-p} we see that 
 the continuous transition line
between that 1-FRSB and the FRSB phases 
displays a point of vertical slope in the 
$(\mu_p,\mu_2)$ plane.
Along the continuous transition line between the 1-FRSB and FRSB phases 
the point of vertical slope is attained for
\begin{equation}
 m^{(\infty)} = \frac{2}{3}\,\frac{p-3}{p-2}
\end{equation}
where 
\begin{eqnarray}
\mu_p^{(\infty)} &=& \frac{(p-2)\,(4 p - 3)^p}{6\,(p-1)(p-3)^{p-1}}\\
\mu_2^{(\infty)} &=& \frac{p\,(p-2)\,(4 p - 3)^2}{12\,(p-3)^2}
\end{eqnarray}
and $q_0=q_1= (p-2)/p$. This point exists for any $p>3$. 

Similarly the point of infinite slope along
the transition line between the 1-FRSB and the 1RSB is attained for 
\begin{equation}
m^{(\infty)} = \frac{2\,(1-y_{\rm max})}{(p-2)\,y_{\rm max}}
\end{equation}
where $y_{\rm max}$ is given by the solution of Eq. (\ref{eq:ymax1rsb}). 
For this value one has 
\begin{eqnarray}
\mu_p^{(\infty)} &=& \frac{(p-2)\,(4 p - 3)^p}{6\,(p-1)(p-3)^{p-1}},\\
\mu_2^{(\infty)} &=& \frac{p^p\, y_{\rm max}(1-y_{\rm max})}{4\,(p-2)^{p-2}}
\end{eqnarray}
and $q_0=0$, $q_1 = (p-2)/p$.
This point exists only for $3< p < 3.5197...$.

\subsection{The Full Replica Symmetry Broken Solution (FRSB)}
For the FRSB solution the order parameter function $q(x)$ is {\em continuous}.
The equations for the FRSB phase are easily obtained from those
of the 1-FRSB by setting $q_0=q_1$ and $m=1$ so that only the continuous part
of the order parameter function survives. In the FRSB phase
the function $x(q)$ is still 
given by Eq. (\ref{eq:xqfrsb}) but with $q_0=q_1$ solution of 
[see Eq. (\ref{eq:rep1frsb})]
\begin{equation}
\label{eq:q0frsb}
\Sigma(q_1) = \frac{1}{(1-q_1)^2}
\end{equation}

\begin{figure}
\includegraphics[scale=1.0]{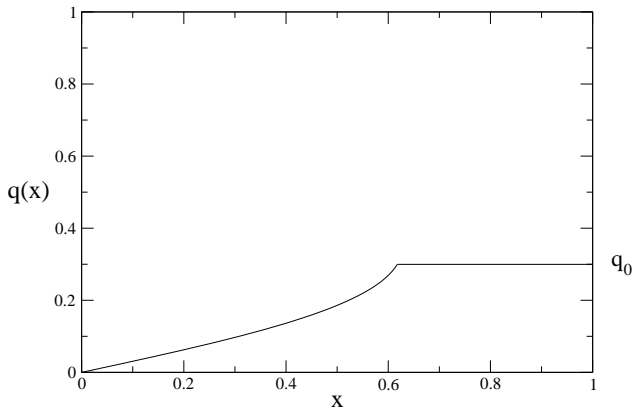}
\caption{Order parameter function $q(x)$ in the FRSB phase. In the figure
         for $p=4$ $\mu_4 = 2$ and  $\mu_2 = 1.5$.
}
\label{fig:qx-2+4-1.5-2}
\end{figure}

The order parameter function $q(x)$ in the FRSB is shown in Figure 
\ref{fig:qx-2+4-1.5-2}.

By defining $\tau = \mu_2 - 1$
from eqs. (\ref{eq:q0frsb}) and (\ref{eq:xqfrsb})
it follows that when the RS instability line $\mu_2=1$ 
is approached from the FRSB side then
\begin{equation}
q_1 \sim \frac{\tau}{2}, \quad \tau\to 0^+
\end{equation}
and 
\begin{equation}
x_c \sim \mu_p\, \frac{(p-1)(p-2)}{2}\, \left(\frac{\tau}{2}\right)^{p-3},
\quad \tau\to 0^+
\end{equation}
It is easy to see that for a generic $p$ the expansion of $q_0$ in powers
of $\tau$ coincide with that of $1-1/\sqrt{1+\tau}$ up to order 
$O(\tau^{p-2})$ not included. For example for $p=4$ one has
\begin{equation}
q_1 \sim \frac{\tau}{2} + \frac{3}{8}(\mu_4 - 1)\tau^2 + O(\tau^3)
\end{equation}
while for $p=5$ 
\begin{equation}
q_1 \sim \frac{\tau}{2} - \frac{3}{8}\tau^2 + 
          \frac{1}{16}(5+4\mu_5)\tau^3 + O(\tau^4)
\end{equation}
and so on.

The transition between the FRSB and RS phases occurs for
$\mu_2=1$ where both $q_1$ and $x_c=x(q_1)$ vanish and the FRSB solution
goes over the RS solution $q=0$. The transition line $(\mu_p,1)$ ends at the 
crossing point with the $m$-line with $m=1$. The transition between the FRSB 
and the RS phases is continuous in the order parameter function and hence 
in the free energy.

\section{Complexity}
\label{complexity}
The 1RSB and 1-FRSB {\it ansatz} both contain the parameter $m$ which
gives the location of the discontinuity in the order parameter
function.  Strictly speaking the replica calculation does not give a
rule to fix it.  Going back to the expression of the moments of the
replica partition function, Eq.  (\ref{eq:znav}), one indeed sees that
the replica calculation requires that for {\em any} overlap matrix
$q_{\alpha\beta}$ the free energy functional must be extremized for
$N\to\infty$ with respect to the {\em elements} of the
matrix. However, it does not say anything about the structure of the
matrix $q_{\alpha\beta}$.  This means that in the 1RSB [and 1-FRSB]
{\it ansatz} the free energy functional $\Phi$ must be extremized with
respect to $q_1$ [and $q(x)$] but not necessarily with respect to $m$,
since it is related to the matrix structure.  This rises the question
of which value of $m$ has to be taken when there exist different
values of $m$, all of which leading to a stable solution. In the
solution discussed so far the value of $m$ yielding the maximum of the
free energy\cite{note4} was chosen.

The free energy functional, Eq.  (\ref{eq:free1}), evaluated on the stable 
saddle point solution gives the free energy of a single pure state.
As a consequence, choosing for $m$ the value which maximizes the free energy
functional is thermodynamically correct, provided that the logarithm of the 
number of different pure states with the same free energy, called 
{\em complexity} or {\em configurational entropy}, 
is not extensive. 
If the configurational entropy
is extensive, it gives a contribution to the thermodynamic free energy
which must be considered when computing the extrema. In other words if
the number of states is extensive the extrema of the thermodynamic free 
energy follow from a balance between the single state free energy 
and the configurational entropy. 
This is what happens in systems with a 1RSB phase, as first noted
in the $p$-spin model,\cite{CriHorSom93,KirWol87} and changes
the condition for fixing the value of $m$.

We shall not give here the details of the direct calculation of the 
complexity for the $2+p$ model, but rather we shall use the shortcut 
of deriving it from a Legendre transform of the replica free energy functional 
with respect to $m$. 

To be more specific the complexity $\Sigma_{\rm lt}$ in the 1RSB and 1-FRSB 
phases is obtained as the Legendre transform of $\beta m \Phi(m)$ where 
$\Phi(m)$ is the replica free energy functional (\ref{eq:free1}) evaluated 
with the 1RSB or 1-FRSB {\it ansatz} keeping
$m$ as a free parameter:
\begin{equation}
\label{eq:LegTr}
\Sigma_{\rm lt}(f) = \max_{m}[\beta m f - \beta m \Phi(m)]
\end{equation}
We shall use for the complexity the notation $\Sigma_{\rm lt}$ to
stress that it is obtained from the Legendre transform, and to
distinguish it from the ``self-energy'' function $\Sigma(x)$ used in
section \ref{static}.  Strictly speaking this is the complexity
density, even if it is customary to call it just complexity.  In the
Legendre transform, Eq.  (\ref{eq:LegTr}), $f$ is the variable
conjugated to $m$
\begin{equation}
f = \frac{\partial m\Phi(m)}{\partial m}
\label{eq:f_m}
\end{equation}
and its value equals the value of the free energy inside a single pure state
 for the given value
of $m$. Introducing this expression into the Legendre transform one gets
the following relation
\begin{equation}
\label{eq:complx}
\Sigma_{\rm lt}(f) = \beta m^2\left. \frac{\partial \Phi(m)}{\partial
m}\right|_{m(f)}
\end{equation}
where $m(f)$ is the value of $m$ found by solving Eq. (\ref{eq:f_m}).

By using the expression (\ref{eq:1frsb-f}) for the functional $G[{\bm q}]$ 
the complexity of the 1-FRSB solutions of the $2+p$ model reads:
\begin{eqnarray}
2\, \Sigma_{\rm lt}(m) &=& -m^2\,\bigl[g(q_1) - g(q_0)\bigr]
     - \ln\left[\frac{\chi(q_1)}{\chi(q_0)}\right] \nonumber\\
&\phantom{=}&
     - m\, \frac{q_1-q_0}{\chi(q_0)}\nonumber 
     +m^2\, (q_1 - q_0)\, \Lambda(q_0)
\end{eqnarray}
where $q_0$ and $q_1$ must be evaluated as function of $\mu_p$, 
$\mu_2$ and $m$ using the saddle point equations (\ref{eq:q11frsb})
and (\ref{eq:rep1frsb}).
Alternatively we can use Eq. (\ref{eq:q11frsb}) to 
eliminate $m$ in favor of $q_1$ so that the expression of the
complexity for the 1-FRSB solutions becomes:
\begin{eqnarray}
\label{eq:cmplx1frsb}
2\Sigma_{\rm lt}(q_1) &=& 
          1
          - \ln\frac{[\Lambda(q_1) - \Lambda(q_0)](1-q_1)^2}{(q_1 - q_0)}
\nonumber\\
&\phantom{=}&
          - \frac{(q_1 - q_0)}{[\Lambda(q_1) - \Lambda(q_0)](1-q_1)^2}
\nonumber\\
&\phantom{=}&
          -\left[\frac{1}{1-q_1} 
                 - (1-q_1)\frac{\Lambda(q_1) - \Lambda(q_0)}{q_1-q_0}\right]^2
\nonumber\\
&\phantom{=}&
\times 
\frac{g(q_1)-g(q_0) - (q_1 - q_0)\Lambda(q_1)}{[\Lambda(q_1) - \Lambda(q_0)]^2}
\end{eqnarray}
where $q_0$ is given by the solution of
\begin{equation}
\label{eq:cmplx-q0}
\mu_2+\mu_p(p-1)q_0^{p-2} = \frac{(1-q_1)^2\,\bigl[\Lambda(q_1) - \Lambda(q_0)\bigr]^2}
                   {(q_1-q_0)^2}
\end{equation}
and $q_1$ is such that $\Phi(q_1) = f$, i.e., $m(q_1) = m(f)$.

The complexity for the 1RSB solution is obtained just setting 
$q_0=0$ into the 1-FRSB complexity [and neglecting Eq. (\ref{eq:cmplx-q0})]. 
A simple check of the 1RSB complexity consists in verifying that
for $\mu_2=0$ one recovers the complexity of
the spherical $p$-spin model.\cite{CriSom95}

By varying $q_1$ one selects 1RSB or 1-FRSB solutions with different
$m$. As a consequence not all values of $q_1$ between $0$ and $1$ are allowed
but only those which lead to stable solutions must be considered. 
This means non-negative
eigenvalues $\Lambda_1^{(1)}$ and $\Lambda_0^{(3)}$ for 1RSB solutions
and non-negative eigenvalue $\Lambda_1^{(1)}$ for 1-FRSB solutions.
The eigenvalue $\Lambda_0^{(3)}$ is identically zero for 1-FRSB solutions.
\begin{figure}
\includegraphics[scale=1.0]{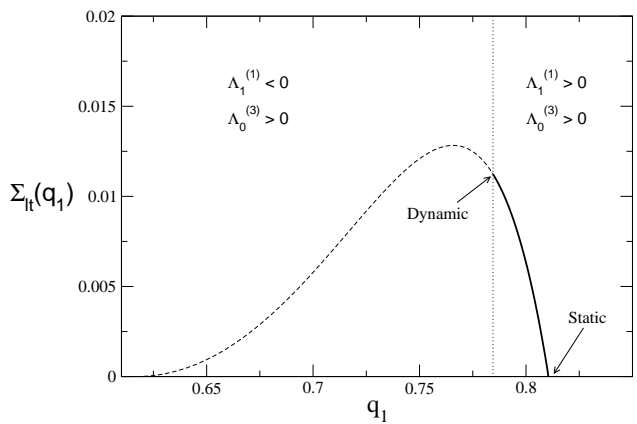}
\caption{$\Sigma_{\rm lt}(q_1)$ as function of $q_1$ in the region where only
        1RSB solutions are stable and have non-negative complexity. 
	Only the part of the curve $\Sigma_{\rm lt}(q_1)\geq 0$ is shown.
	The thick line shows the relevant part of the curve. The dashed line
	correspond to unstable solutions. The point marked ``Dynamic'' 
	is where $\Lambda_1^{(1)}$ vanishes and corresponds to the solution
	of largest complexity.
	The figure is for the
	$2+4$ model with $\mu_2 = 3$ and $\mu_4 = 10$.
        }
\label{fig:cmplx_24_3-10}
\end{figure}
The requirement that only solutions with non-negative $\Lambda_1^{(1)}$
are physically acceptable is also know as the 
Plefka's criterion.\cite{Plefka,CriLeuRiz03} Here it
comes out naturally from the stability analysis of the replica saddle point,
however it can be shown to have a more general validity.

The complexity $\Sigma_{\rm lt}$ is the logarithm of the number
of states of given free energy, divided by the system size $N$.
 It is, therefore, clear that in the thermodynamic limit only solutions 
with a non-negative complexity must be considered. All others will be 
exponentially depressed and hence are irrelevant.

The static solution discussed in previous Sections was obtained by imposing
$\partial \Phi(m) /\partial m = 0$. The complexity $\Sigma_{\rm lt}$ 
is consequently zero for the static solution, and the number of ground states 
is not extensive.

\begin{figure}
\includegraphics[scale=1.0]{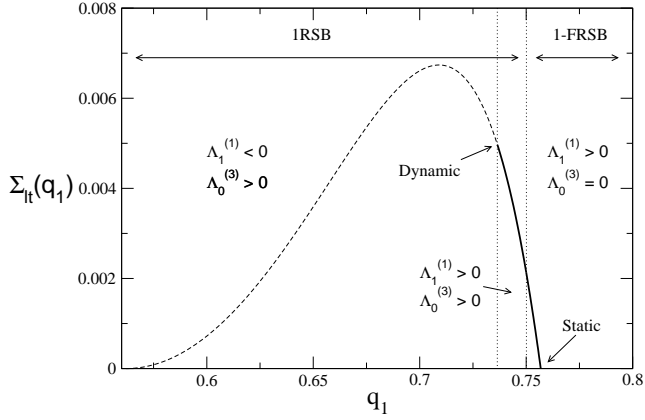}
\caption{$\Sigma_{\rm lt}(q_1)$ as function of $q_1$ in the region where both
        1RSB and 1-FRSB solutions are stable and have non-negative complexity. 
	We have indicated explicitly which part of the curve corresponds to 
	each solutions.
	Only the part of the curve $\Sigma_{\rm lt}(q_1)\geq 0$ is shown.
	The thick line shows the relevant part of the curve. The dashed line
	correspond to unstable solutions. The point marked ``Dynamic'' 
	is where $\Lambda_1^{(1)}$ vanishes and corresponds to the solution
	of largest complexity.
	The figure is for the 
	$2+4$ model with $\mu_2 = 3$ and $\mu_4 = 7$.
        }
\label{fig:cmplx_24_3-7}
\end{figure}

\begin{figure}
\includegraphics[scale=1.0]{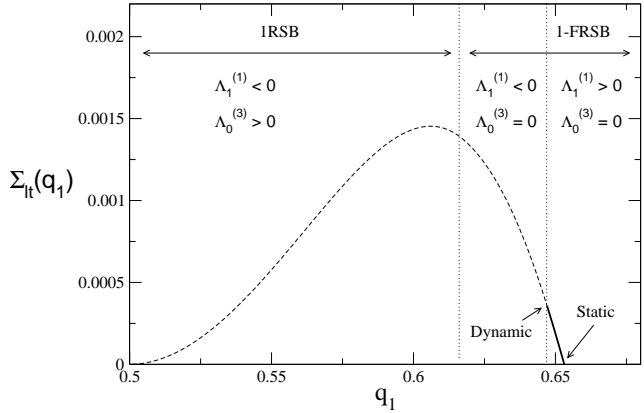}
\caption{$\Sigma_{\rm lt}(q_1)$ as function of $q_1$ in the region where 
        1RSB and 1-FRSB solutions have non-negative complexity but only the
	latter are stable. 
	We have indicated explicitly which part of the curve corresponds to 
	each solutions.
	Only the part of the curve $\Sigma_{\rm lt}(q_1)\geq 0$ is shown.
	The thick line shows the relevant part of the curve. The dashed line
	correspond to unstable solutions. The point marked ``Dynamic'' 
	is where $\Lambda_1^{(1)}$ vanishes and corresponds to the solution
	of largest complexity.
	The figure is for the 
	$2+4$ model with $\mu_2 = 3$ and $\mu_4 = 4$.
        }
\label{fig:cmplx_24_3-4}
\end{figure}

The solution with the largest complexity, of both 1RSB or 1-FRSB type, 
is the one for which $\Lambda_1^{(1)}$
vanishes, i.e.,\cite{CriSom95,CriLeuRiz03} 
\begin{equation}
\label{eq:margc}
\Sigma(q_1) = \frac{1}{(1-q_1)^2}.
\end{equation}
In Figures \ref{fig:cmplx_24_3-10}, \ref{fig:cmplx_24_3-7} and 
\ref{fig:cmplx_24_3-4} we show the behavior of $\Sigma_{\rm lt}(q_1)$
in the three relevant regions where (i) only 1RSB solutions have non-negative 
complexity and are stable, (ii)  both 1RSB and 1-FRSB solutions have 
non-negative complexity and are stable and (iii) both 1RSB and 1-FRSB solutions
have non-negative complexity but only the latter are stable.

The condition of maximal complexity (\ref{eq:margc}) is known as the 
``marginal condition'' since for $\Lambda_1^{(1)}=0$ the saddle point
is marginally stable.

In the relaxation dynamics the eigenvalue $\Lambda_1^{(1)}$ is related to 
the decay 
of the two-times correlation function to the ``intermediate'' value $q_1$,
and hence the marginal condition comes naturally in as the condition 
for critical decay.\cite{CriHorSom93,CiuCri00} 
For this reason the solution of maximal complexity is also called the 
``dynamic solution'' as opposed to the ``static solution'' discussed so far
which, on the contrary, has vanishing complexity. 

In the FRSB phase $\Lambda_1^{(1)}$ is identically zero and 
the two solutions, static and dynamic, coincide.


\section{The Dynamic Solution}
\label{dynamics}
In this paper we shall not give here the full derivation of the dynamic 
solution, and of the marginal condition (\ref{eq:margc}),
starting from the relaxation dynamic equations
but rather we shall rely on the fact that 
the dynamic solution can be obtained from the replica calculation just 
using the marginal condition instead of stationarity of the replica free 
energy functional with respect to $m$.\cite{CriHorSom93,CriSom95}
It can be shown that this shortcut
applies also to the 1-FRSB solution.\cite{CriLeu05}

The static and dynamic solutions differ only for what concern the 1RSB and 
1-FRSB phases, therefore here we shall only discuss shortly the main 
differences in the phase diagram which follows from 
the 1RSB and 1-FRSB dynamic solutions.

\subsection{The Dynamic 1RSB Solution}
The equations of the dynamic 1RSB solution are given by 
Eq. (\ref{eq:q11rsb}) with 
$q_0=0$ and by the marginal condition (\ref{eq:margc}).
Solving these equations for $(\mu_p,\mu_2)$ and using 
$q_1$ as a free parameter one gets the parametric equations of the 
dynamical 1RSB $m$-lines
\begin{eqnarray}
\label{eq:mup1rsbdy}
\mu_p &=& \frac{m}{(p-2)\,q_1^{p-3}(1-q_1)^2 (1-q_1 + mq_1)}\\
\label{eq:mu21rsbdy}
\mu_2 &=& \frac{(p-2)(1-q_1) - mq_1}{(p-2)\,(1-q_1)^2 (1-q_1 + mq_1)}
\end{eqnarray}
For any $0\leq m\leq 1$ 
the maximum allowable value of $q_1$ is fixed 
by the requirement that $\mu_2 \geq 0$, while the 
minimum by the requirement that the eigenvalue 
$\Lambda_3^{(0)}$, Eq. (\ref{eq:rep1rsb}) with $q_0=0$, be non-negative. 
In the dynamic approach this eigenvalue controls the long time relaxation of
the two-times correlation function\cite{CriHorSom93} 
and hence must be non-negative.
A straightforward calculation shows that 
\begin{equation}
\label{eq:dy1rsblim}
\frac{p-3}{p-3+m}\leq q_1\leq \frac{p-2}{p-2+m}, \qquad 
\end{equation}
with $0\leq m\leq 1$.

\subsubsection{The Dynamic Transition Line Between the Paramagnet and the 1RSB-Glass Phase}
The dynamic transition line between the RS and the 1RSB phases is given
by the dynamic 1RSB $m$-line with $m=1$. Inserting $m=1$ into 
eqs. (\ref{eq:mup1rsbdy})-(\ref{eq:mu21rsbdy}) 
one obtains the parametric equations of the transition line
\begin{eqnarray}
\mu_p &=& \frac{1}{(p-2)\,q_1^{p-3}(1-q_1)^2}\\
\mu_2 &=& \frac{(p-2) - (p-1)q_1}{(p-2)\,(1-q_1)^2}
\end{eqnarray}
with $(p-3)/(p-2)\leq q_1\leq (p-2)/(p-1)$. In the $(\mu_p,\mu_2)$ plane 
the line begins on the $\mu_p$
axis at the point
\begin{equation}
\mu_p = \frac{(p-1)^{p-1}}{(p-2)^{p-2}}, \quad \mu_2 = 0
\end{equation}
and goes up till the point 
\begin{equation}
\label{eq:ep1rsbdy}
\mu_p = \frac{(p-2)^{p-2}}{(p-3)^{p-3}}, \quad \mu_2 = 1.
\end{equation}
where $\Lambda_3^{(0)}$ vanishes. The transition between the RS and 
the dynamic 1RSB phase is discontinuous in the 
order parameter $q_1$ since it jumps from zero, on the RS side, to a finite
value on the $m$-line with $m=1$.

The dynamic 1RSB phase is bounded by the critical line of equation
$\Lambda_3^{(0)}=0$ which
marks the transition between the dynamic 1RSB and the dynamic 
1-FRSB phases.
The explicit form of the equation of this transition line 
is obtained by
inserting $q_1=(p-3)/(p-3+m)$, see Eq. (\ref{eq:dy1rsblim}), 
into the equations of the dynamic 1RSB $m$-line and reads
\begin{eqnarray}
\label{eq:mup1rsbc}
\mu_p &=& \frac{(p-3+m)^p}{(p-2)^2\, (p-3)^{p-3}\, m^2}\\
\label{eq:mu21rsbc}
\mu_2 &=& \frac{(p-3+m)^2}{(p-2)^2\, m^2}
\end{eqnarray}
with $0\leq m\leq 1$. 

\begin{figure}
\includegraphics[scale=1.0]{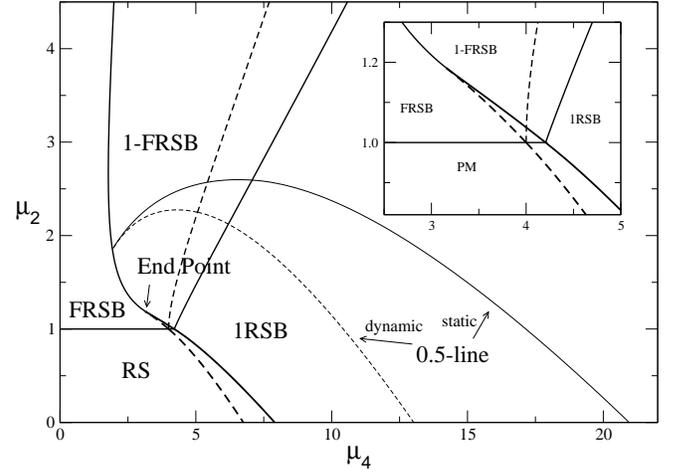}
\caption{Phase diagram $\mu_2$-$\mu_4$. Counterclockwise the RS,
  1RSB, 1-FRSB and
  FRSB phases are plotted, separated by the {\it static} phase
  transition lines (full curves).  The PM/1RSB and the 1RSB/1-FRSB
  transitions also occur in the {\em dynamics} and the relative lines
  are drawn as dashed curves.  
  Their
  continuation as FRSB/1-FRSB transition lines are the dynamic
  (dashed) and static (full) $m=1$-lines, computed in the 1-FRSB 
  {\it ansatz}. They merge at the ``End
  Point'' (see inset).  For a comparison, we also plot the dynamic and
  static $m$-lines with $m=0.5$.  They merge on the FRSB/1-FRSB phase
  transition line above the end point. As $m$ decreases from one, the
  whole {\em continuous} FRSB/1-FRSB line is covered.  Inset: the
  discontinuous transitions FRSB/1-FRSB ($\mu_2>1$) and PM/1RSB
  ($\mu_2<1$).  }
\label{fig:phdi.2+4}
\end{figure}

As expected, the dynamic transition lines do not coincide 
with the static ones but, in the $(\mu_p,\mu_2)$ plane, they are displaced toward lower 
values of $\mu_p$ 
with respect to the corresponding static transition lines, 
see Figure \ref{fig:phdi.2+4}.


\subsection{The Dynamic 1-FRSB Solution}
The equations of the dynamic 1-FRSB solution are given by the saddle point 
equations (\ref{eq:qx1frsb}) and (\ref{eq:q11frsb}) and by the marginal
condition (\ref{eq:margc}). As a consequence, the parametric equations
of the dynamic 1-FRSB $m$-lines are still 
(\ref{eq:mup1frsb})-(\ref{eq:mu21frsb}) but with
the value of $y$ given by
\begin{equation}
\label{eq:y1frsbdy}
y = \frac{1 - (p-1)\,t^{p-1} + (p-2)\, t^{p-2}}
         {p-2 - (p-1)\, t + t^{p-1}}
\end{equation}
where $0\leq t=q_1/q_0\leq 1$.
The continuous part of the order parameter function is given by 
Eq. (\ref{eq:xqfrsb}) with [see Eq. (\ref{eq:auxy})], 
\begin{equation}
q_0 = t\,q_1 = \frac{t\,(1-y)}{1-y+my\,(1-t)}
\end{equation}

The dynamic 1-FRSB $m$-line are drawn in the $(\mu_p,\mu_2)$ plane 
by fixing the value of $m$ and varying $t$ from $0$ to $1$.

%


\subsubsection{The Dynamic Transition Line between the 1RSB and the 1-FRSB 
Amorphous Phases}

By setting $t=0$ into the equations of the dynamic 1-FRSB $m$-lines 
and varying $m$ from $1$ to $0$ one recovers the critical line
$q_0=0$ and $\Lambda_3^{(0)}=0$ which marks the boundary with the 
dynamic 1RSB phase.
Indeed for $t=0$
Eq. (\ref{eq:y1frsbdy}) yields $y=1/(p-2)$ so that 
eqs. (\ref{eq:mup1frsb})-(\ref{eq:mu21frsb}) reduce to the parametric
equations (\ref{eq:mup1rsbc})-(\ref{eq:mu21rsbc}) of the critical line.
Moreover on this line $q_1=(p-3) / (p-3+m)$, the same value found from the
dynamic 1RSB solution, therefore 
as it happens for the static solution 
the dynamic 1-FRSB $m$-lines and the dynamic 1RSB $m$-lines with the same $m$
match continuously on the critical line $\Lambda_3^{(0)}=0$ [and $q_0=0$].
This transition is continuous since $q_0$ goes 
to zero as the transition line is approached from the 1-FRSB side while
$q_1$ is continuous through the line.

The dynamic 1-FRSB $m$-line with $m=1$, continuation of the 1RSB $m$-line with
$m=1$ into the 1-FRSB phase, marks the boundary between
the 1-FRSB and FRSB phases. Along this line the order parameter is 
discontinuous since $q_1$ jumps from zero in the FRSB phase to a non-zero
value on the line. The discontinuity occurs at $m=1$ so that the free energy 
remains continuous despite the jump in the order parameter.
The dynamic 1-FRSB $m$-line with $m=1$ starts from the end point 
(\ref{eq:ep1rsbdy}) of the dynamic 1RSB $m$-line with $m=1$  and 
stops at the same end point (\ref{eq:mupcrit})-(\ref{eq:mu2crit}) of the
static 1-FRSB $m$-line with $m=1$. From this point the transition 
between the 1-FRSB phase and the FRSB phase occurs continuously in the
order parameter, i.e., with $q_1-q_0\to 0$ as the transition line
is approached form the 1-FRSB side (see Fig. \ref {fig:phdi.2+4}).

The continuous transition between the 1-FRSB and FRSB phases occurs along the
critical line obtained by setting $t=1$ into 
eqs. (\ref{eq:mup1frsb})-(\ref{eq:mu21frsb}) and varying $m$ from $1$ to $0$.
From Eq. (\ref{eq:y1frsbdy}) it follows that 
\begin{equation}
y = 1 - \frac{p-3}{2}\,(1-t) + O((1-t)^2), \qquad t\to 1^-
\end{equation}
so that the end point of the dynamic 1-FRSB $m$-line coincides
with the end points of the static 1-FRSB $m$-line for any $m$, and not only for
$m=1$. Therefore, the dynamic and the static continuous critical lines
between the 1-FRSB and the FRSB phases coincide.
Indeed,  when studying the static solution we have seen that 
along the continuous transition line between the
1-FRSB and the FRSB solutions the eigenvalue $\Lambda_1^{(1)}$ vanishes 
so that the difference between the two solutions disappears. On this line
both solutions have zero complexity and it remains equal to zero in 
the whole FRSB phase.

In Figure \ref{fig:phdi.2+4} we show the full phase diagram of the 
spherical $2+p$ spin glass model in the $(\mu_p,\mu_2)$ space with
both the static and the dynamic critical lines.

\begin{figure}
\includegraphics[width=.49\textwidth]{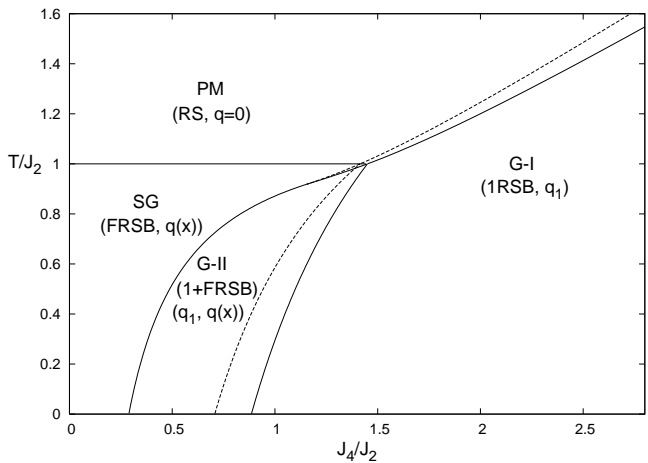}
\caption{Phase diagram $T$-$J_4$ in $J_2$ units. Clockwise a paramagnetic phase
(PM),
a glassy phase (G-I) described by means of a 1RSB {\em ansatz},
another glassy phase represented by a 1-FRSB {\em ansatz} and a 
purely spin-glass
phase (SG) computed in the FRSB {\em ansatz} occur.
The transitions to and from the two glass-like phases take place
 both as dynamic (dotted lines) 
and as static (full lines). In temperature 
the dynamic transition lines are always above the relative static lines.} 
\label{fig:phdi.TJ.2+4}
\end{figure}

By noticing that both $\mu_p$ and $\mu_2$ are proportional to $\beta^2$ we see
that the discontinuous dynamic transition occurs at a temperature
higher than that of the equivalent static transition, as can be clearly seen
from Figure \ref{fig:phdi.TJ.2+4} where 
the phase diagram in the
$T/J_2=1/\sqrt{\mu_2}$ and $J_p/J_2=\sqrt{2\mu_p/(p\mu_2)}$ plane
is shown.

\section{Conclusions}

In this paper we have provided a detailed study of the phase space of
the spherical $2+p$ spin glass model using the static approach
of Ref. [\onlinecite{CriSom92}] which employs the replica method to evaluate
the disorder-averaged logarithm of the partition function. By performing the
Legendre transform of the replica free energy functional we have 
defined the complexity function that, whenever it is extensive, 
counts the number of equivalent different metastable states. This allowed
us to discuss the dynamic solution as the solution which maximizes the
complexity. In both solutions, i.e., static and dynamic, 
the model displays four different phases, characterized by different
replica symmetry breaking schemes, in which the system can find
itself as the thermodynamic parameters and the intensity of the
interaction is changed. One of the nice feature of this model is that
it can be completely solved even in the phase described by a full replica
symmetry breaking. To our knowledge this is the first example of an
analytical FRSB solution.

The main result of the present study is the existence of
two phases exhibiting the qualitative features of glassy materials.
One is the 1RSB solution, displayed by those systems that are
considered as valid mean-field models for the glass state.  Even
though no connection with the microscopic constituents of a real glass
former can be set, the collection of spins interacting through a
$p$-body quenched disordered bound behaves very much like, e.g., the
set of SiO$_2$ molecules in a window glass.
The single breaking occurring in the 1RSB solution corresponds, in a
dynamic interpretation \cite{SompPRL81,DeDom82, stocstab} to a time-scale
bifurcation between the fast processes in a real amorphous material
(the $\beta$-processes) and the slow processes ($\alpha$) responsible
for the structural relaxation.

Another phase emerges in the study of the spherical $2+p$ spin glass
model.  Something not occurring in any Ising spin glass
model.\cite{note12} In a whole region of the phase space the stable
phase is, indeed, described by means of an overlap function $q(x)$
that is continuous up to a certain value $q_0$ and then displays a
step, as in aforementioned 1RSB solution. We call it the one step-full
RSB solution (1FRSB).  Exploiting the static-dynamic analogy once
again, in this phase, in the relaxation towards equilibrium, a first
time-scales bifurcation takes place (``$\alpha-\beta$ bifurcation'')
just as above, but it is no more unique.  As the time goes by a
continuous set of further bifurcations starts to occur between slow
and even slower processes, as in the case of a proper
spin-glass.\cite{REV} In the continuous part, any kind of similarity
between (ultrametrically organized) states is allowed but above $q_0$
the hierarchy ends in only one extra possible value: the self-overlap,
or Edwards-Anderson order parameter $q_1$.

The stability of the 1-FRSB solution in the replica space is not
limited to a single, self-consistent, choice of the order parameter
(in particular the point discontinuity can change in a certain
interval) and this implies that in each point of the phase diagram
belonging to this phase there will be an extensive number of metastable
states, having free energies higher than the equilibrium free energy.
In order for this phase to appear a strong enough couple interaction
must be present (that is the source of the continuous, spin-glass like
contribution) but the $p$-body interaction must have a broader
distribution of intensities than the $2$-body ($J_p>J_2$).

If this new phase can be considered as a glassy phase different from
the 1RSB one, the phase diagram that we have computed describes an
amorphous-amorphous transition, with the second glass having a much
more complicated structure outside the single valleys (for very long
time-scales in the dynamic language).  Whether there is a
correspondence with the amorphous-amorphous transition between
hard-core repulsive and attractive glassy
colloids\cite{DawetAl01,Zac01,Sciorti02,Eckert02, Pham02} and at which
level the analogy can be set is yet to be clarified and further
investigation of the dynamic properties is needed in order to make a
link between the interplay of $2$-body and $p$-body interaction in
spherical spin glasses and the role of repulsive and attractive
potentials in colloids undergoing kinetic arrest.


\appendix
\section{The Parisi equation and the Sommers-Dupont formalism}
\label{app:SomDup}
The Parisi equation for the $2+p$ spin glass model is more easily obtained
starting from the free energy functional in the replica space written as
\begin{eqnarray}
\label{eq:freepar}
 \beta f[{\bm q}, {\bm\lambda}] &=& - s(\infty) 
                 - \lim_{n\to 0}\frac{1}{2n} 
                     \sum_{\alpha\beta}\, g(q_{\alpha\beta})
\nonumber\\
&\phantom{=}& 
                + \lim_{n\to 0}\frac{1}{2n} 
            \sum_{\alpha\beta}\, \lambda_{\alpha\beta}\,q_{\alpha\beta}
\\
&\phantom{=}& 
  -\lim_{n\to 0}\frac{1}{n} \ln{\mbox Tr}_{\sigma}\, \exp\left(
     \frac{1}{2}\sum_{\alpha\beta}\, \lambda_{\alpha\beta}\,
                   \sigma^{\alpha}\sigma^{\beta}
                 \right)
\nonumber
\end{eqnarray}
where the matrix $\lambda_{\alpha\beta}$ 
is the Lagrange multiplier associated with the replica overlap matrix 
$q_{\alpha\beta}$, see Eq. (\ref{eq:qab}). In particular the diagonal element
$\lambda_{\alpha\alpha}=\overline{\lambda}$ is the Lagrange multiplier that 
enforces the spherical constraint $q_{\alpha\alpha}=1$.
Stationary of $f[{\bm q},{\bm \lambda}]$ with respect to variations of 
$\lambda_{\alpha\beta}$ and $q_{\alpha\beta}$ leads to the self-consistent
equations $(\alpha\not=\beta$)
\begin{eqnarray}
\lambda_{\alpha\beta} &=& \Lambda(q_{\alpha\beta}) \\
q_{\alpha\beta} &=& \frac
           {
     {\mbox Tr}_{\sigma}\,\sigma^{\alpha}\sigma^{\beta}\,
         \exp{
          \left(\sum_{\alpha\beta}\,
             \lambda_{\alpha\beta}\,\sigma^{\alpha}\sigma^{\beta}
          \right)}
           }
          {
     {\mbox Tr}_{\sigma}\,
         \exp{
          \left(\sum_{\alpha\beta}\,
             \lambda_{\alpha\beta}\,\sigma^{\alpha}\sigma^{\beta}
          \right)}
          }
\end{eqnarray}

By applying the Parisi's replica symmetry breaking scheme an infinite number 
of times and introducing
the functions $\lambda(x)$ and $q(x)$, $0\leq x\leq 1$, one for each matrix,
the free energy functional (\ref{eq:freepar}) for the spherical model can be
written as
\begin{eqnarray}
\beta f &=& - s(\infty) 
          - \frac{1}{2}\bigl[
              g(1) - \int_{0}^{1}\,dx\,g(q(x))
                       \bigr] 
\nonumber\\
&\phantom{=}&
          + \frac{1}{2}\bigl[
                  \overline{\lambda}
                 - \int_{0}^{1}\,dx\,\lambda(x)\,q(x)
                       \bigr] 
    - \frac{1}{2}\ln\left(\frac{2\pi}{\lambda(1) - \overline{\lambda}}\right) 
\nonumber\\
&\phantom{=}&
          - \int_{-\infty}^{+\infty}\, \frac{dy}{\sqrt{2\pi\lambda(0)}}\,
             \exp\left(-\frac{y^2}{2\lambda(0)}\right)\,\phi(0,y)
\end{eqnarray}            
where $\phi(0,y)$ is the solution evaluated for $x=0$ of the Parisi equation
\begin{equation}
\label{eq:pareq}
\dot{\phi}(x,y) = -\frac{1}{2}\dot{\lambda}(x)\left[
                \phi''(x,y) + x\,\phi'(x,y)^2\right]
\end{equation}
with the boundary condition
\begin{equation}
\label{eq:initc}
 \phi(1,y) = \frac{1}{2}\,\frac{y^2}{\lambda(1) - \overline{\lambda}}
\end{equation}
In writing the Parisi equation we have used the standard notation
in which a dot ``$\dot{\phi}$'' denotes the derivative with respect 
to $x$ 
while the prime ``$\phi'$'' the derivative with respect to $y$.

The advantage of this equation is that it can be solved numerically
without specifying a-priori the form of $q(x)$ and $\lambda(x)$.  The
first problem one is facing when solving the Parisi equation is that
the functional (\ref{eq:freepar}) must be extremized over all possible
solutions of the Parisi equations, which can be numerically
uncomfortable.  This, however, can be overcome using the
Sommers-Dupont formalism.\cite{SomDup84}  The idea is to introduce a different
functional whose value at the stationary point coincides with the
extrema of the free energy functional (\ref{eq:freepar}) over all
possible solutions of the Parisi equations.

This is easily achieved by introducing the Parisi equation into the functional
via the Lagrange multiplier $P(x,y)$. The new functional is hence
\begin{eqnarray}
\label{eq:freesom}
\beta f_{v} &=& \beta f 
              + \int_{-\infty}^{+\infty}\,dy\, P(1,y)\left[
           \phi(1,y) - \frac{1}{2}\frac{y^2}{\lambda(1) - \overline{\lambda}}
                     \right]
\nonumber\\
&\phantom{=}& 
          - \int_{0}^{1}\, dx\,\int_{-\infty}^{+\infty}\, dy P(x,y)\Bigl[
           \dot{\phi}(x,y) 
\nonumber\\
&\phantom{=}& 
\phantom{xxx}
         + \frac{1}{2}\dot{\lambda}(x)\left[
                \phi''(x,y) + x\,\phi'(x,y)^2\Bigr]
	   \right]
\end{eqnarray}
where $\beta f$ is the free energy functional (\ref{eq:freepar}).

The functional $\beta f_v$ is stationary with respect to variations of
$P(x,y)$, $P(1,y)$, $\phi(x,y)$, $\phi(0,y)$, the order parameter
functions $q(x)$ and $\lambda(x)$ and $\overline{\lambda}$. 
Stationarity with respect to $P(x,y)$ and $P(1,y)$ just gives back equations
(\ref{eq:pareq}) and (\ref{eq:initc}), while stationarity with respect to
$\phi(x,y)$ and $\phi(0,y)$ leads to the differential equation for
$P(x,y)$:
\begin{equation}
\label{eq:pxy}
\dot{P}(x,y) =\frac{1}{2}\dot{\lambda}(x)\left[
                P''(x,y) -2x\,\bigl[P(x,y)\,m(x,y)\bigr]'\right]
\end{equation}
where $m(x,y) = \phi'(x,y)$,
with the boundary condition
\begin{eqnarray}
\label{eq:pinitc}
 P(0,y) &=& \frac{1}{\sqrt{2\pi\lambda(0)}}\,
             \exp\left(-\frac{y^2}{2\lambda(0)}\right)
\nonumber \\
&=& \delta(y), \quad\ {\rm as}\ \lambda(0)\to 0
\end{eqnarray}

It can be shown that $P(x,y)$ is the probability distribution
of the local field $y$ at the scale $q(x)$.

Finally stationarity with respect to $q(x)$, $\lambda(x)$ and 
$\overline{\lambda}$ gives 
\begin{equation}
\label{eq:lamx}
\lambda(x) = \Lambda(q(x))
\end{equation}
\begin{equation}
\label{eq:qxsom}
q(x) = \int_{-\infty}^{+\infty}\, dy\, P(x,y)\, m(x,y)^2
\end{equation}
and
\begin{equation}
\label{eq:overl}
\lambda(1) - \overline{\lambda} = \frac{1}{1-q(1)}
\end{equation}

From Eq. (\ref{eq:qxsom}) and the identification of $P(x,y)$ with the
local field distribution it follows that $m(x,y)$ represents 
the local magnetization at scale $q(x)$ in presence of the local field $y$.
It obeys the differential equation
\begin{equation}
\label{eq:mxy}
\dot{m}(x,y) = - \frac{1}{2}\dot{\lambda}(x)\left[
                m''(x,y) + 2x\,m(x,y)\,m'(x,y)\right]
\end{equation}
with initial condition
\begin{equation}
\label{eq:minitc}
 m(1,y) = [1-q(1)]\, y
\end{equation}

\begin{figure}
\includegraphics[scale=1.0]{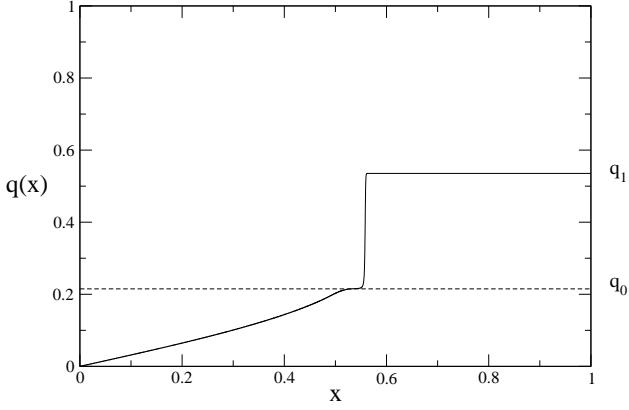}
\caption{The numerical order parameter function $q(x)$ for the $2+4$ model
         with $\mu_2=2$ and $\mu_4=3$. The 1-FRSB solution is clearly 
	 seen. 
}
\label{fig:qx-2+4-2.0-3.0-n}
\end{figure}

\begin{figure}
\includegraphics[scale=1.0]{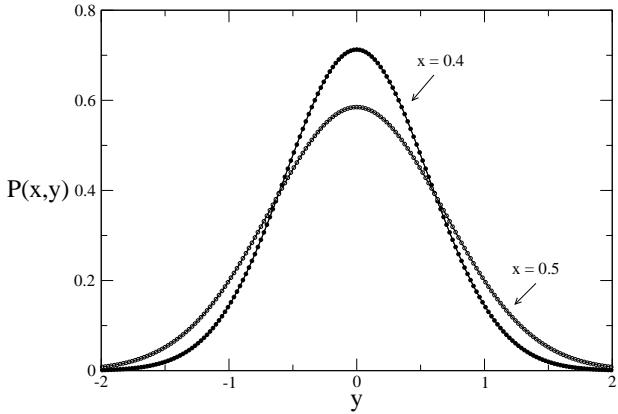}
\caption{Field probability distribution $p(x,y)$ for the $2+4$ model
with $\mu_2 = 2$ and $\mu_4 = 3$ evaluated from the numerical resolution
of the Parisi-Sommers-Dupont equations at 
$x = 0.4$ and $x = 0.5$.
The underlying full curve (barely visible) is the analytic 
solution given by Eqs.  (\ref{pxy}), (\ref{eq:sigma-pxy}).}
\label{fig:pxy}
\end{figure}

The partial differential equations (\ref{eq:pxy}) and (\ref{eq:mxy}) can 
be solved numerically using the pseudo-spectral method developed in
Refs. [\onlinecite{CriRiz02,CLP02}].
In Figure \ref{fig:qx-2+4-2.0-3.0-n} we show the order parameter function
$q(x)$ found solving the equations for the $2+4$ model with $\mu_2=2$ and 
$\mu_4=3$. The 1-FRSB structure is clearly seen.
In Fig. \ref{fig:pxy} both the numerical and the analytical solutions for
$P(x,y)$ are displayed.

\section{Solution of the Parisi-Sommers-Dupont equations}
\label{app:SomDup-sol}
It is not difficult to write the Parisi-Sommers-Dupont partial differential 
equations for the 1-FRSB case, cfr. Ref. [\onlinecite{CriLeuParRiz04}].
With obvious notation, the Parisi-Sommers-Dupont functional with the
1-FRSB {\it ansatz} reads
\begin{eqnarray}
\label{eq:freesom1frsb}
\beta f_{v} &=& \beta f 
              + \int_{-\infty}^{+\infty}\,dy\, P(m,y)\Big[
           \phi(m,y) 
\nonumber\\
&\phantom{=}& 
\phantom{xxx}
        - \frac{1}{2}\frac{y^2}
       {\lambda_1 - \overline{\lambda} - m\,(\lambda_1 - \lambda_0)}
                     \Bigr]
\nonumber\\
&\phantom{=}& 
          - \int_{0}^{m}\, dx\,\int_{-\infty}^{+\infty}\, dy P(x,y)\Bigl[
           \dot{\phi}(x,y) 
\nonumber\\
&\phantom{=}& 
\phantom{xxx}
         + \frac{1}{2}\dot{\lambda}(x)\left[
                \phi''(x,y) + x\,\phi'(x,y)^2\Bigr]
	   \right]
\end{eqnarray}
where $\beta f$ is the free energy functional
\begin{eqnarray}
\beta f &=& - s(\infty) 
\nonumber\\
&\phantom{=}&
          - \frac{1}{2}\bigl[
              g(1) - (1-m)\,g(q_1) - \int_{0}^{m}\,dx\,g(q(x))
                       \bigr] 
\nonumber\\
&\phantom{=}&
          + \frac{1}{2}\bigl[
                  \overline{\lambda}
                 - (1-m)\,\lambda_1 q_1
                 - \int_{0}^{m}\,dx\,\lambda(x)\,q(x)
                       \bigr] 
\nonumber\\
&\phantom{=}&
    - \frac{1}{2}\ln\left(\frac{2\pi}{\lambda(1) - \overline{\lambda}}\right) 
\nonumber\\
&\phantom{=}&
    - \frac{1}{2m}\ln\left(
        \frac{\lambda_1 - \overline{\lambda}}
             {\lambda_1 - \overline{\lambda} - m\,(\lambda_1 - \lambda_0)}
	     \right)
\nonumber\\
&\phantom{=}&
          - \int_{-\infty}^{+\infty}\, \frac{dy}{\sqrt{2\pi\lambda(0)}}\,
             \exp\left(-\frac{y^2}{2\lambda(0)}\right)\,\phi(0,y)
\end{eqnarray}            

Stationarity of (\ref{eq:freesom1frsb}) with respect to variations of
$P(x,y)$ and $P(m,x)$ yields the Parisi equation (\ref{eq:pareq}) with
$x$ restricted to the interval $[0,m]$ and initial condition
\begin{equation}
\label{eq:initc1rsb}
\phi(m,y) =         - \frac{1}{2}\frac{y^2}
       {\lambda_1 - \overline{\lambda} - m\,(\lambda_1 - \lambda_0)}
\end{equation}
Similarly stationarity with respect to variations of 
$\phi(x,y)$, $\phi(0,y)$, $q(x)$, $\lambda(x)$ for $0\leq x\leq m$
gives again eqs. (\ref{eq:pxy}), (\ref{eq:pinitc}), (\ref{eq:lamx}) and
(\ref{eq:qxsom}), while stationarity with respect to $\overline{\lambda}$
leads back to eq (\ref{eq:overl}) with $\lambda(1)$ replaced by $\lambda_1$.

For the 1-FRSB {\it ansatz} there are three more equations. The 
first two follow from stationarity with respect to $q_1$, $\lambda_1$ and read
\begin{equation}
\lambda_1 = \Lambda(q_1)
\end{equation}
\begin{equation}
\lambda_1 - \lambda_0 = \frac{q_1 - q_0}{(1-q_1)[1 - q_1 - m\,(q_1 - q_0)]}
\end{equation}
Finally stationarity with respect to
$m$, the discontinuity point in the
order parameter function, leads again to equation (\ref{eq:m1frsb}).
It is not difficult to recognize in these equations the saddle point 
equations for $q_0$, $q_1$ and $m$ derived for the 1-FRSB phase.

For the spherical model the Parisi-Sommers-Dupont equations can be solved 
analytically. Indeed defining 
\begin{equation}
\label{eq:effe}
F(x) = \lambda_1 - \overline{\lambda} - m\,(\lambda_1 - \lambda_0) 
       - \int_{x}^{m}\, dx'\, x'\, \dot{\lambda}(x')
\end{equation}
it is easy to verify that the solution reads
\begin{equation}
\phi(x,y) = \frac{1}{2}\left[
            \frac{y^2}{F(x)} 
          + \int_{x}^{m}\,dx'\,\frac{\dot{\lambda}(x')}{F(x')}
                       \right]
\end{equation}
\begin{equation}
m(x,y) = \frac{y}{F(x)}
\label{mxy}
\end{equation}
and
\begin{equation}
\label{eq:locfd}
P(x,y) = \frac{1}{\sqrt{2\pi\sigma(x)^2}}\, 
               \exp\left(-\frac{y^2}{2\sigma(x)^2}\right)
\label{pxy}
\end{equation}
where
\begin{equation}
\sigma(x)^2 = F(x)^2\,\int_{0}^{x}\,dx' \frac{\dot{\lambda}(x')}{F(x')^2}
\end{equation}
Inserting Eqs. (\ref{mxy})-(\ref{pxy})  into Eq. (\ref{eq:qxsom}) one gets 
\begin{equation}
\label{eq:qxf}
q(x) = \int_{0}^{x}\, dx'\, \frac{\dot{\lambda}(x')}{F(x')^2} 
\end{equation}
Taking the derivative of eqs. (\ref{eq:effe}) and (\ref{eq:qxf}) with respect 
to $x$ it is easy to show that 
\begin{equation}
\frac{d F}{F^2} = x\, dq
\end{equation}
which integrated yields the equality $\chi(x) = 1/F(x)$. Thus Eq. 
(\ref{eq:qxf}) becomes 
\begin{equation}
q(x) = \int_{0}^{x}\, dx'\, \dot{\lambda}(x')\,\chi(x')^2
\end{equation}
Inverting this relation one gets back Eq. (\ref{eq:qx1frsb}).
We note that changing the integration variable from $x$ to $q$ in Eq. 
(\ref{eq:qxf}) one can show that $F^2(q) = \Sigma(q)$ so that the variance 
$\sigma(x)^2$ of the local field distribution (\ref{eq:locfd}) 
can be written as
\begin{equation}
\label{eq:sigma-pxy}
\sigma(x)^2 = \Sigma(q(x))\, q(x)
\end{equation}

This completes the solution of the Parisi-Sommers-Dupont equations for the
spherical $2+p$ spin glass model in the 1-FRSB phase. The FRSB solution
can be obtained taking $q_0\to q_1$ and $m\to 1$.


\section{The spherical $s+p$ spin glass model}
\label{app:s+p}

The spherical $s+p$ spin glass model is the generalization of $2+p$
models in which the two-spin interaction is replaced by a $s$-body
spin interaction:
\begin{equation}
\label{eq:ham-s+p}
{\cal H} = \sum_{i_1<\ldots <i_s}^{1,N}J^{(s)}_{i_1\ldots i_s}
            \sigma_{i_1}\cdots\sigma_{i_s}
           +\sum_{i_1<\ldots <i_p}^{1,N}J^{(p)}_{i_1\ldots i_p}
            \sigma_{i_1}\cdots\sigma_{i_p}
\end{equation}
In the following we shall assume that $s<p$, even if
the Hamiltonian is trivially invariant under the exchange of $s$ and $p$.

The study of the $s+p$ model follows closely that of the $2+p$ with the
replacement of $g(x)$ by [see Eq. (\ref{eq:free3})]
\begin{equation}
g(x) = \frac{\mu_s}{s} x^s + \frac{\mu_p}{p} x^p.
\end{equation}
As it happens for the $2+p$ models the RS solution with $q\not=0$ is
unstable, however, at difference with these, the RS solution with $q=0$  is
stable everywhere in the $(\mu_p,\mu_s)$ plane
since for $q=0$ the relevant eigenvalue $\Lambda_1$, eq.(\ref{eq:rs-stab}),
is identically equal to one for any $s>2$.

For $\mu_p$ and/or $\mu_2$ large enough a thermodynamically more
favorable 1RSB solution appears. This solution has $q_0=0$ and $q_1$
given by the saddle point equation (\ref{eq:q11rsb}) [with $q_0=0$].
For the static solution the value of $m$ is fixed by Eq. (\ref{eq:m1rsb}),
or equivalently by Eq. (\ref{eq:z1rsb}), which follows from stationarity of
the free energy functional with respect to variations of $m$.
These equations can be solved for any $s$ and $p$ with the help of the
CS $z$-function and one obtains the
parametric equations of the static $m$-lines:
\begin{eqnarray}
\label{eq:mup1rsb-s+p}
\mu_p &=& \frac{p}{(p-s)}\,
          \frac{(1-y+m\,y)^p}{m^2 y(1-y)^{p-2}}\,
          \frac{[2 -s\,z(y)]}{2}
\\
\label{eq:mus1rsb-s+p}
\mu_s &=& \frac{s}{(p-s)}\,
          \frac{(1-y+m\,y)^s}{m^2 y(1-y)^{s-2}}\,
          \frac{[p\,z(y) - 2]}{2}
\end{eqnarray}
where $y = (1-q_1) / (1-q_1 + m\,q_1)$.
Notice that, as expected, the expressions are symmetric under the exchange
of $s$ and $p$.

The 1RSB solution never becomes unstable since for $s>2$ the
eigenvalue $\Lambda_0^{(3)}$, Eq. (\ref{eq:rep1rsb-1}), evaluated at $q_0=0$
is always positive, and the eigenvalue $\Lambda_1^{(1)}$,
Eq. (\ref{eq:rep1rsb}), remains positive for any $\mu_p\geq 0$.

As a consequence the limit on $y$ are given by the conditions
\begin{eqnarray}
\mu_p = 0\ &\Rightarrow&\ z(y_{\mu_p}) = \frac{2}{s} \\
\mu_s = 0\ &\Rightarrow&\ z(y_{\mu_s}) = \frac{2}{p}
\end{eqnarray}
The CS $z$-function is an increasing function of $y$ therefore
if $s<p$ then $y_{\mu_p}<y_{\mu_s}$. For $s=3$ and $p=4$ we have
\begin{equation}
y_{\mu_4} = 0.195478...,\quad  y_{\mu_3} =  0.354993...
\end{equation}

To find the dynamic transition line between the 1RSB and the RS phase
equation (\ref{eq:z1rsb}) must be replaced by
the marginal condition (\ref{eq:margc}). Using $q_1$ as the independent
variable it is easy to derive the parametric
equations for the dynamic $m$-lines:
\begin{eqnarray}
\label{eq:mup1rsb-s+p-d}
\mu_p &=& \frac{1}{(p-s)}\,
          \frac{(s-2+m)\,q_1 - (s-2)}
               {q_1^{p-2}\,(1-q_1)^2\,(1-q_1+m\,q_1)}
\\
\label{eq:mus1rsb-s+p-d}
\mu_s &=& \frac{1}{(p-s)}\,
          \frac{(p-2) - (p-2+m)\,q_1}
               {q_1^{s-2}\,(1-q_1)^2\,(1-q_1+m\,q_1)}
\end{eqnarray}
The range of $q_1$ is fixed by the requirement that both $\mu_p$ and $\mu_s$
be non-negative. This yields the boundary values
\begin{eqnarray}
\mu_p = 0\ &\Rightarrow&\ q_1 = \frac{s-2}{s-2+m} \\
\mu_s = 0\ &\Rightarrow&\ q_1 = \frac{p-2}{p-2+m}
\end{eqnarray}

\begin{figure}
\includegraphics[scale=1.0]{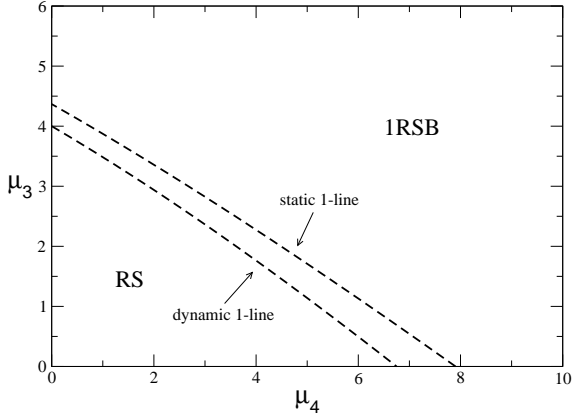}
\caption{The phase diagram of the $3+4$ model in the $(\mu_3,\mu_4)$
         plane. The thick dashed lines are the
         $m$-line with $m=1$, i.e. represent 
 the discontinuous transition between
         the RS and the 1RSB phases.
        }
\label{fig:sp341rsb}
\end{figure}

In Figure \ref{fig:sp341rsb} we show the phase diagram for the $s+p$ model
with $s>2$ and $p>s$. In the figure $s=3$ and $p=4$, however any choice of
$s>2$ and $p> s$ leads to a qualitatively similar phase diagram.


\section{The RSB Solutions}
\label{app:stabil}
In this Appendix we show that the spherical $2+p$ spin glass model 
admits only solutions of  1RSB, FRSB or 1-FRSB type.
The procedure is the same as that of Appendix 2 of 
Ref. [\onlinecite{CriSom92}].

We start from the free energy functional for the $R$-RSB {\it ansatz} 
[see Eq. (\ref{eq:free-rsb})]
\begin{eqnarray}
\frac{2}{n}\, G[{\bm q}] &=& 
            \int_{0}^{1} dq~ x(q)\, \Lambda(q) 
\nonumber\\
&\phantom{=}&
            + \int_{0}^{q_R} \frac{dq}{\int_{q}^{1} dq'\, x(q')}
            + \ln\left(1 - q_R\right)
\label{eq:arpl-f}
\end{eqnarray}
where 
\begin{equation}
\label{eq:axqr}
x(q) = p_0 + \sum_{r=0}^{R} (p_{r+1} - p_r)\, \theta(q - q_r).
\end{equation}

\begin{figure}
\includegraphics[scale=1.0]{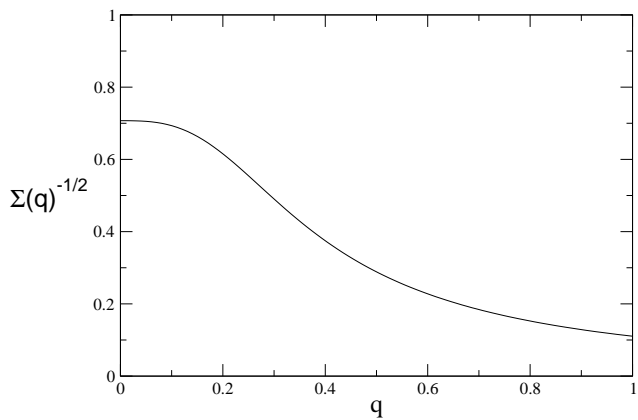}
\caption{Typical behavior of 
$1/\sqrt{\Sigma(q)}$ as function of $q$ for $p>3$.}
\label{fig:sigma-q-1sqr}
\end{figure}

The saddle point equations are obtained by varying
the above functional with respect to $x(q)$:
\begin{equation}
\frac{2}{n}\,\delta G[{\bm q}] = \int_{0}^{1}\,dq\, F(q)\,\delta x(q)
\end{equation}
where
\begin{equation}
F(q) = \Lambda(q) - \int_{0}^{q} \frac{dq'}
                  {\left[\int_{q'}^{1} dq''\, x(q'')\right]^2}
\end{equation}
and
\begin{eqnarray}
 \delta x(q) &=& \sum_{r=0}^{R}\,(\delta p_{r+1} - \delta p_r)\,
                     \theta(q - q_r)
\nonumber \\
&\phantom{=}& 
    - \sum_{r=0}^{R}\,(p_{r+1} - p_r)\,\delta(q-q_r)\,\delta q_r
\end{eqnarray}
By requiring stationarity of $G[{\bm q}]$ with respect to the $q_r$ and the
$p_r$ one gets, respectively
\begin{equation}
\label{eq:fq0}
F(q_r) = 0, \qquad r = 0,\ldots, R
\end{equation}
\begin{equation}
\label{eq:mst}
\int_{q_{r-1}}^{q_r}\, dq\, F(q) = 0, \qquad r = 1,\ldots, R
\end{equation}
The function $F(q)$ is continuous, thus Eq. (\ref{eq:mst}) implies
that between any two successive $q_r$ there must be at least two
extrema of $F(q)$.  If we denote these by $q^*$, then the extremal
condition $F'(q^*)=0$ implies that
\begin{equation}
\label{eq:der}
\int_{q^*}^{1}\, dq\, x(q) = \frac{1}{\sqrt{\Sigma(q^*)}} 
\end{equation}
The left hand side of this equation is a concave function, i.e. with a 
negative second derivative, since $x(q)$ is not decreasing with $q$.
For the $2+p$ model $\Sigma(q) = \mu_2 + \mu_p (p-1) q^{p-2}$ so that 
the right hand side is concave for small $q$, provided $p>3$, and convex
for large $q$, see Figure \ref{fig:sigma-q-1sqr}.
For $p=3$ the right hand side is convex.

As a consequence Eq. (\ref{eq:der}) admits at most two solutions
and hence only one step of replica symmetry breaking is possible.
Moreover it must be $q_0=0$. 
Indeed, from Eq. (\ref{eq:fq0}) and from the fact that in 
absence of external field $F(0)=0$ it follows that
\begin{equation}
F(0) = F(q_0) = F(q_1)= 0
\end{equation}
which would imply the presence of at least three extrema, which is not 
possible.

Up to this point the conclusions do not differ much from those found for
the $p$-spin model. Here however the presence of a concave part in the
right hand side of Eq. (\ref{eq:der}) for $p>3$ makes possible
different solutions.

Equations (\ref{eq:fq0}) and (\ref{eq:mst}) can be solved by 
a continuous replica symmetry breaking solution with 
$F(q)=0$ in a given range of $q$ since in this case both equations
would be identically valid. For this solution Eq. (\ref{eq:der})
must also be identically valid and this can only be true for 
$0\leq q\leq q_0$, where 
$q_0$ is the value of $q$ for which
$1/\sqrt{\Sigma(q)}$ changes concavity,
where both sides of the equation are concave function of $q$.
Indeed if Eq. (\ref{eq:der}) were valid also for values of $q$ where the 
right hand side is convex this would imply that $x(q)$ is a 
decreasing function of $q$, which is not possible since $dx(q)/dq$ is the 
probability density of the overlaps.
For the same reason 
for $p=3$ a continuous replica symmetry breaking solution is not allowed 
since in this case the right hand side of Eq. (\ref{eq:der}) is
purely convex. The same applies to the $s+p$ models with $s>2$,
so that also these models admits only a 1RSB phase.

The possibility of a continuous replica symmetry breaking solution for 
$q\leq q_0$ does not rule out the presence of discrete replica breakings
with $q_r>q_0$. These additional discrete breakings must satisfy
eqs. (\ref{eq:fq0}) and (\ref{eq:mst}). Therefore
since the right hand side of Eq. (\ref{eq:der}) is convex for $q>q_0$ 
arguments similar to those which leaded to the 
conclusion that for the $2+3$ model only 1RSB solutions are possible
show that at most only one more discrete break 
with $q_1>q_0$ is possible.

No other possible non trivial solutions exist for the spherical $2+p$ spin 
glass model so we conclude that the model admits only
solutions of 1RSB, 1-FRSB or FRSB type.



\begin{thebibliography}{99}
\bibitem{note21}
  The replica trick uses the identity 
  $\ln \overline{x} = \lim_{n\to 0} \ln\overline{x^n}/n$ and hence 
  $\Phi(n)$ should be known for a continuous set of values of $n$ and not 
  just for integers.
  Consequence of this is that the limit $\lim_{n\to 0} \Phi(n)$ is not
  the analytic continuation of $\Phi(n)$ down to $n=0$.
\bibitem{note7}
  The calculation of the eigenvalues of the quadratic form 
  (\protect\ref{eq:stabil}) for 1RSB solution of the $2+p$ model 
   does not differ much from that of the spherical $p$-spin model. 
   As a consequence to help the reader interested into the details of the 
   derivation we shall use for the eigenvalues 
   the same notation as Ref. [\protect\onlinecite{CriSom92}].
\bibitem{note4}
For finite $n>1$ the saddle point calculation requires that $G[{\bm q}]$
be maximal at the stationary point. However the dimension 
$n(n-1)/2$ of the ${\bm q}$-space becomes negative for $n<1$
so that $G[{\bm q}]$ should be minimized in the limit $n\to 0$.
Consistently the free energy functional is maximized for $n\to 0$.
\bibitem{note5} 
The lowest value of the order parameter matrix 
$q_{\alpha\beta}$ is the minimum possible overlap between replicas.
It is always zero in absence of an external field.
\bibitem{note1}
  The most general form of  the order parameter function in the 
  1RSB phase is $q(x) = q_0 \theta(m-x) + q_1\theta(x-m)$.
  However, in  absence of external fields, $q_0=0$.
\bibitem{note8}
  To maintain the notation $q_1$ for the overlap associated to the 1RSB part 
  of the order parameter function 
  we use the notation $q_0$ for the largest value of the continuous part $q(x)$
  of the order parameter function. This notation should not be
  confusing since $q(0)=0$.
\bibitem{note6}
In the expansion given in Ref.[\protect\onlinecite{CriLeu04}] the exponent 
of the third term is erroneously given as $x^7$.
\end{thebibliography}
\end{document}